\documentclass[12pt]{article}

\usepackage{amssymb}
\usepackage{amsmath}
\usepackage{amsfonts}

\newcommand{\R}{\mathbb{R}}
\newcommand{\C}{\mathbb{C}}

\newcommand{\be}{\begin{equation}}
\newcommand{\ee}{\end{equation}}
\newcommand{\bea}{\begin{eqnarray}}
\newcommand{\eea}{\end{eqnarray}}
\newcommand{\nn}{\nonumber}
\newcommand{\kt}{\rangle}
\newcommand{\br}{\langle}

\newcommand{\cum}{\mbox{\scriptsize${\cal M}$}}
\newcommand{\ed}{\end{document}}

\begin{document}

\title{Conserved Current Densities, Localization
Probabilities, and a New Global Gauge Symmetry of
Klein-Gordon~Fields}
\author{A.~Mostafazadeh\thanks{Corresponding author, E-mail address:
amostafazadeh@ku.edu.tr}~and  F.~Zamani\thanks{E-mail address:
zamani@iasbs.ac.ir} \\ \\
$^*$~Department of Mathematics, Ko\c{c} University,
Rumelifeneri Yolu, \\ 34450 Sariyer, Istanbul, Turkey \\
$ ^\dagger$~Department of Physics, Institute for Advanced Studies
in Basic \\ Sciences, 45195-159 Zanjan, Iran}
\date{ }
\maketitle

\begin{abstract} For free Klein-Gordon fields, we construct a
one-parameter family of conserved current densities $J_a^\mu$,
with $a\in(-1,1)$, and use the latter to yield a manifestly
covariant expression for the most general positive-definite and
Lorentz-invariant inner product on the space of solutions of the
Klein-Gordon equation. Employing a recently developed method of
constructing the Hilbert space and observables for Klein-Gordon
fields, we then obtain the probability current density ${\cal
J}_a^\mu$ for the localization of a Klein-Gordon field in space.
We show that in the nonrelativistic limit both $J_a^\mu$ and
${\cal J}_a^\mu$ tend to the probability current density for the
localization of a nonrelativistic free particle in space, but that
unlike $J_a^\mu$ the current density ${\cal J}_a^\mu$ is neither
covariant nor conserved. Because the total probability may be
obtained by integrating either of these two current densities over
the whole space, the conservation of the total probability may be
viewed as a consequence of the local conservation of $J_a^\mu$.
The latter is a manifestation of a previously unnoticed global
gauge symmetry of the Klein-Gordon fields. The corresponding gauge
group is $U(1)$ if the parameter $a$ is rational. It is the
multiplicative group of positive real numbers if $a$ is
irrational. We also discuss an extension of our results to
Klein-Gordon fields minimally coupled to an electromagnetic field.
\end{abstract}
~~~~~~~~PACS numbers: 11.30.-j, 11.40.-q, 03.65.Pm, 98.80.Qc

\baselineskip=24pt

\textheight = 22cm \topskip = -1cm \topmargin = -1cm

\section{Introduction}

The demand for devising a probabilistic interpretation for
Klein-Gordon fields is among the oldest problems of modern
physics. Though this problem was never fully resolved, it provided
the motivation for some of the most important developments of the
twentieth-century theoretical physics. The most notable of these
are the discovery of the Dirac equation and the advent of the
method of second-quantization which eventually led to the
formulation of the quantum field theories. The latter, in turn,
provided the grounds for completely disregarding the original
problem of finding a probabilistic interpretation for Klein-Gordon
fields as there were sufficient evidence that the first-quantized
(scalar) field theories involved certain inconsistencies such as
the Klein paradox \cite{holstein}.\footnote{See however
\cite{ghose}.} There were also some general arguments suggesting
that the localization of bosonic fields in space was not possible
\cite{peierls}. These led to the consensus that the correct
physical picture was provided by second-quantized field theories
and that one could safely neglect the above-mentioned problem, for
it only arose if one dealt with the first-quantized fields.

Though this point of view is almost universally accepted, the fact
that a logical transition from nonrelativistic to relativistic
quantum mechanics requires studying the first-quantized fields
compels one to appeal to the conventional interpretation of
Klein-Gordon fields in terms of the Klein-Gordon current density:
    \be
    J_{\rm KG}^\mu = ig\,\psi(x^0,\vec x)^*
    \stackrel{\leftrightarrow}{\partial^\mu}\psi(x^0,\vec x).
    \label{kgc}
    \ee
Here $g\in\R^+$ is a constant, $\psi$ is a solution of the
Klein-Gordon equation
    \be
    [\partial_\mu\partial^\mu-\cum^2]\psi(x^0,\vec x)=0,
    \label{kg}
    \ee
$\cum:=mc/\hbar$ is the inverse of the Compton's wave length, $m$
is the mass of $\psi$,
$\partial_\mu\partial^\mu=\eta^{\mu\nu}\partial_\mu\partial_\nu$,
$\eta^{\mu\nu}$ are components of the inverse of the Minkowski
metric $(\eta_{\mu\nu})$ with signature $(-1,1,1,1)$, and for any
pair of Klein-Gordon fields $\psi_1$ and $\psi_2$, $\psi_1
\stackrel{\leftrightarrow}{\partial^\mu} \psi_2:=
\psi_1\partial^\mu\psi_2- (\partial^\mu\psi_1)\psi_2$.

As it is discussed in most textbooks on relativistic quantum
mechanics, $J_{\rm KG}^\mu$ is a conserved four-vector current
density, i.e., it is a four-vector field satisfying the continuity
equation
    \be
    \partial_\mu J_{\rm KG}^\mu=0.
    \label{conti-eq}
    \ee
Therefore, $J_{\rm KG}^0$ may be used to defined a conserved
quantity, namely
    \be
    Q:= \int_{\R^3} d^3\vec x~J_{\rm KG}^0(x^0,\vec x).
    \label{q}
    \end{equation}

The fact that $Q$ (respectively $J_{\rm KG}^0$) takes positive as
well as negative values does not allow one to identify it with a
probability (respectively probability density). Instead, one
identifies $Q$ with the electric charge of the field and views
$J_{\rm KG}^\mu$ as the corresponding four-vector charge density
\cite{greiner}. In this way the continuity
equation~(\ref{conti-eq}) provides a differential manifestation of
the electric charge conservation.

It can be shown that $Q$ takes positive values for positive-energy
Klein-Gordon fields and that one can define a Lorentz-invariant
positive-definite inner product on the set of positive-energy
fields. This is the basis of the point of view according to which
one identifies the physical Hilbert space with the subspace of
positive-energy fields \cite{wald-gr}. This approach however fails
for the cases that the Klein-Gordon field is subject to a
time-dependent background field, for in this case the notion of a
positive-energy Klein-Gordon field is ill-defined. Even in the
absence of time-dependent background fields, the above restriction
to the positive-energy fields limits the choice of possible
observables to those that do not mix positive- and negative-energy
fields. There are also well-known (and related) problems regarding
the violation of causality \cite{causality}.\footnote{See however
\cite{zuben,barat-kimball} and references therein.} The safest way
out of all these difficulties seems to be a total abandonment of
the first-quantized field theories as viable physical theories
\cite{ryder}.

The interest in the issue of finding a probabilistic
interpretation for first-quantized Klein-Gordon fields was revived
in the 1960s within the context of canonical quantum gravity.
There it emerged as a fundamental obstacle in developing a quantum
theory of gravity \cite{bryce-67}. This time neither Dirac's trick
of considering an associated first order field equation nor the
application of the method of second-quantization could be applied
satisfactorily \cite{qg-review}. It was also not possible to
define a subset of positive-energy solutions which would serve as
the underlying vector space for the `physical Hilbert space' of
the theory. It was then necessary to deal with the above-described
problems with first-quantized fields directly.

The lack of a probabilistic interpretation for canonical quantum
gravity and its simplified version known as quantum cosmology is
widely referred to as the {\em Hilbert-space problem},
\cite{qg-review}. This terminology reflects the view that the
problem of devising a probabilistic interpretation for the wave
functions appearing in these theories (namely the Wheeler-DeWitt
fields) is equivalent to constructing a Hilbert space to which
these fields belong. In this way one can identify the observables
of the theory with Hermitian operators acting in this Hilbert
space and utilize Born's probabilistic interpretation of quantum
mechanics.

The Klein-Gordon charge density may be used to define an inner
product on the space of solutions of the Klein-Gordon equation.
This is known as the Klein-Gordon inner product:
    \be
    (\psi_1,\psi_2)_{_{\rm KG}}= i g\int_{\R^3} d^3\vec x~
    [\psi_1(x^0,\vec x)^*\dot\psi_2(x^0,\vec x)-
    \dot\psi_1(x^0,\vec x)^*
    \psi_2(x^0,\vec x)],
    \label{kgi}
    \ee
where $\psi_1$ and $\psi_2$ are Klein-Gordon fields, $g$ is a
nonzero positive real constant, and an overdot stands for a
$x^0$-derivative, i.e., $\partial_0$. Clearly, $(\psi,\psi)_{_{\rm
KG}}=\int d^3\vec x J_{\rm KG}^0(x^0,\vec x)=Q$. As $Q$ may be
negative, the Klein-Gordon inner product is indefinite
\cite{indefinite-m}. Therefore, endowing the (vector) space ${\cal
V}$ of solutions of the Klein-Gordon equation with the
Klein-Gordon inner product does not produce a genuine inner
product space. One may pursue the approach of the
indefinite-metric quantum theories \cite{indefinite} and identify
the subspace ${\cal V}_+$ of positive-energy solutions as the
physical space of state vectors. Restricting the Klein-Gordon
inner product to this subspace one obtains a (definite) inner
product space that can be extended to a separable Hilbert space
${\cal H}_+$ through Cauchy completion \cite{reed-simon}. This is
the basis of developing quantum field theories in curved
background spacetimes \cite{wald-qft}.

In order to ensure that the right-hand side of (\ref{kgi}) is
convergent, it is sufficient to assume that for all $x^0\in\R$,
the functions $\psi(x^0),\dot\psi(x^0):\R^3\to\C$ defined by
    \[\psi(x^0)(\vec x):=\psi(x^0,\vec x),~~~~~~~~~~~
    \dot\psi(x^0)(\vec x):=\partial_0\psi(x^0,\vec x),\]
are square-integrable, i.e., $\psi(x^0),\dot\psi(x^0)\in
L^2(\R^3)$. This is an assumption that we shall make throughout
this paper. It is supported by the fact that $\psi$ tends to a
solution of the free Schr\"odinger equation in the nonrelativistic
limit ($c\to\infty$).

We can respectively express the Klein-Gordon equation~(\ref{kg}),
the space ${\cal V}$ of its solutions, and the Klein-Gordon inner
product (\ref{kgi}) in terms of the functions $\psi(x^0)$ and
$\dot\psi(x^0)$ according to
    \bea
    &&\ddot\psi(x^0)+D\psi(x^0)=0,
    \label{kg-0}\\
    &&{\cal V}=\left\{\left.\psi:\R\to L^2(\R^3)~\right|
    \ddot\psi(x^0)+D\psi(x^0)=0\right\},
    \label{v}\\
    &&(\psi_1,\psi_2)_{_{\rm KG}}= i g\left[\br\psi_1(x^0)|
    \dot\psi_2(x^0)\kt-\br\dot\psi_1(x^0)|\psi_2(x^0)\kt\right],
    \label{kgi-0}
    \eea
where $D:L^2(\R^3)\to L^2(\R^3)$ is the operator defined by
    \be
    (D\phi)(\vec x):=(-\nabla^2+\cum^2)\phi(\vec x)~~~~~~~~~~~~~~
    \forall\phi\in L^2(\R^3),
    \label{D}
    \ee
$\psi_1,\psi_2\in{\cal V}$ are arbitrary, and $\br\cdot|\cdot\kt$
stands for the inner product of $L^2(\R^3)$.

Recently, it has been noticed that one can devise a systematic
method of endowing ${\cal V}$ with a positive-definite inner
product \cite{cqg,ap,p57}. In particular, one may define a
positive-definite and relativistically invariant inner product on
${\cal V}$, namely
    \be
    (\psi_1,\psi_2):= \frac{1}{2\cum}\,
    \left[\br\psi_1(x^0)|D^{1/2}\psi_2(x^0)\kt
    +\br\dot\psi_1(x^0)|D^{-1/2}\dot\psi_2(x^0)\kt\right].
    \label{inn1}
    \ee
Note that $D$ and consequently $D^{-1}$ are positive-definite
operators (Hermitian operators with strictly positive spectra) and
$D^{\pm 1/2}$ is the unique positive square root of $D^{\pm 1}$.
In addition to being positive-definite and relativistically
invariant, the inner product~(\ref{inn1}) is a conserved quantity
in the sense that the $x^0$-derivative of the right-hand side of
(\ref{inn1}) vanishes. As explained in Ref.~\cite{ap}, this is
required to make the inner product~(\ref{inn1}) well-defined.

The expression (\ref{inn1}) was originally obtained in \cite{cqg}
using the results of the theory of pseudo-Hermitian operators
\cite{ph}. To the best of our knowledge, the existence of a
positive-definite and relativistically invariant inner product for
Klein-Gordon fields was originally pointed out by the authors of
Refs.~\cite{halliwell-ortiz} and \cite{woodard}. Though these
authors pursue completely different approaches\footnote{The
analysis of \cite{halliwell-ortiz} involves the study of a certain
Green's function for the Klein-Gordon equation, whereas that of
\cite{woodard} employs the idea of gauge-fixing the inner product
of the auxiliary Hilbert space obtained in the quantization of the
classical relativistic particle within the framework of Dirac's
method of constraint quantization. A detailed application of the
latter idea for Klein-Gordon fields is given in
\cite{hartle-marolf,halliwell-thorwart,zuben}. Its direct
extension is the method of refined algebraic quantization also
known as the group-averaging \cite{induced}. For a brief review
see \cite{embacher}.}, they arrive at an expression which is an
alternative form of the inner product (\ref{inn1}). The advantage
of the approach of \cite{cqg} is that not only it yields an
explicit expression for a positive-definite and relativistically
invariant inner product, but it also allows for a construction of
all such inner products: The most general positive-definite,
relativistically invariant, and conserved inner product on the
space ${\cal V}$ of solutions of the Klein-Gordon equation
(\ref{kg-0}) is given by \cite{cqg,ap}
    \bea
    (\psi_1,\psi_2)_a&=&\frac{\kappa}{2\cum}\,
    \left\{\br\psi_1(x^0)|D^{1/2}\psi_2(x^0)\kt
    +\br\dot\psi_1(x^0)|D^{-1/2}\dot\psi_2(x^0)\kt+\right.\nn\\
    &&\hspace{1.5cm}\left. ia\left[\br\psi_1(x^0)|
    \dot\psi_2(x^0)\kt-\br\dot\psi_1(x^0)|
    \psi_2(x^0)\kt\right]\right\},
    \label{gen-i}
    \eea
where $\kappa\in\R^+$ and $a\in(-1,1)$ are arbitrary constants. If
we set $g=1/(2\cum)$ in (\ref{kgi-0}), we can express
(\ref{gen-i}) in the form
    \be
    (\psi_1,\psi_2)_a=\kappa[(\psi_1,\psi_2)+a
    (\psi_1,\psi_2)_{_{\rm KG}}].
    \label{gen-i-0}
    \ee
Note that $\kappa$ is an unimportant multiplicative constant that
can be absorbed in the definition of the Klein-Gordon fields
$\psi_1$ and $\psi_2$.

The existence of the positive-definite inner products
(\ref{gen-i-0}) and their relativistic invariance and conservation
raise the natural question whether there is a conserved
four-vector current density associated with these inner products.
The main purpose of the present article is to show that such a
current density exists. In fact, we will construct a one-parameter
family $J^\mu_a$ (with $a\in(-1,1)$) of current densities and show
by direct computation that not only they transform as vector
fields but they also satisfy the continuity equation and yield the
well-known Schr\"odinger probability current density in
nonrelativistic limit. Furthermore, the current density $J^\mu_a$
may be used to yield a manifestly covariant expression for the
inner products~(\ref{gen-i}). Perhaps more importantly, its local
conservation law is linked with the conservation of the total
probability of the localization of the field in space, on the one
hand, and a previously unnoticed Abelian global gauge symmetry of
the Klein-Gordon equation, on the other hand.

The organization of the article is as follows. In Section~2, we
outline a derivation of the current densities $J_a^\mu$ and
explore their properties. In Section~3, we derive the probability
current density for the localization of a Klein-Gordon field in
space and show how it relates to the current densities $J_a^\mu$.
In Section~4, we study the underlying gauge symmetry associated
with the conservation of $J_a^\mu$. In Section~5, we discuss a
generalization of our results to Klein-Gordon fields interacting
with a background electromagnetic field. In Section~6 we present
our concluding remarks. The appendices include some useful
calculations that are, however, not of primary interest.

\section{Derivation and Properties of $J_a^\mu$}

Let $\psi\in{\cal V}$ be a Klein-Gordon field. Then in view of
(\ref{gen-i}), we have
    \bea
    (\psi,\psi)_a&=&\frac{\kappa}{2\cum}\,
    \int_{\R^3}d^3\vec x
    \left\{\psi(x^0,\vec x)^*\hat D^{1/2}\psi(x^0,\vec x)
    +\dot\psi(x^0,\vec x)^*\hat D^{-1/2}\dot\psi(x^0,\vec x)+
    \right.\nn\\
    &&\hspace{2.5cm}\left. ia\left[\psi(x^0,\vec x)^*
    \dot\psi(x^0,\vec x)-\dot\psi(x^0,\vec x)^*
    \psi(x^0,\vec x)\right]\right\},
    \label{gen-i-exp}
    \eea
with $\hat D:=-\nabla^2+\cum^2$. Now, using the analogy with
nonrelativistic quantum mechanics, we define the current density
$J_a^0$ associated with $\psi$ as the integrand in
(\ref{gen-i-exp}). That is
    \be
    J_a^0(x):=\frac{\kappa}{2\cum}\left\{\psi(x)^*\hat D^{1/2}\psi(x)
    +\dot\psi(x)^*\hat D^{-1/2}\dot\psi(x)+
    ia\left[\psi(x)^*\dot\psi(x)-\dot\psi(x)^*
    \psi(x)\right]\right\},
    \label{j-zero}
    \ee
where we have set $x:=(x^0,\vec x)$.

In order to obtain the spatial components $J_a^i$ (with
$i\in\{1,2,3\}$) of $J_a^\mu$, we follow the procedure outlined in
Ref.~\cite{rh}. Namely, we perform an infinitesimal Lorentz boost
transformation that changes the reference frame to the one moving
with a velocity $\vec v$. That is we consider
    \be
    x^0\to {x'}^0=x^0-\vec \beta\cdot\vec x,~~~~~~~~~~
    \vec x\to\vec x'=\vec x-\vec \beta x^0,
    \label{boost}
    \ee
where $\vec\beta:=\vec v/c$, and we ignore second and higher order
terms in powers of the components of $\vec\beta$. Assuming that
$J_a^\mu$ is indeed a four-vector field, we obtain the following
transformation rule for $J_a^0$.
    \be
    J_a^0(x)\to{J'}_a^0(x')=J_a^0(x)-\vec\beta\cdot\vec J_a(x).
    \label{boost-j}
    \ee
Next, we recall that we can use (\ref{j-zero}) to read off the
expression for ${J'}_a^0(x')$, namely
    \be
    {J'}_a^0(x'):=\frac{\kappa}{2\cum}\left\{\psi'(x')^*
    \hat {D'}^{1/2}\psi'(x')
    +\dot\psi'(x')^*\hat {D'}^{-1/2}\dot\psi'(x')+
    ia\left[\psi'(x')^*\dot\psi'(x')-\dot\psi'(x')^*
    \psi'(x')\right]\right\},
    \label{j-zero-prime}
    \ee
where $x':=({x'}^0,\vec x')$, $\hat D'=-{\nabla'}^2+\cum^2$ and
$\dot\psi'(x'):=\partial\psi'(x')/\partial {x'}^0$. This reduces
the determination of $\vec J_a$ to expressing the right-hand side
of (\ref{j-zero-prime}) in terms of the quantities associated with
the original (unprimed) frame and comparing the resulting
expression with (\ref{boost-j}).

It is obvious that $\psi$ is a scalar field;
    \be
    \psi'(x')=\psi(x).
    \label{scalar}
    \ee
A less obvious fact is that $D^{-1/2}\dot\psi$ is also a scalar
field. This can be directly checked by performing an infinitesimal
Lorentz transformation as we demonstrate in Appendix~A.
Alternatively, we may appeal to the observation that the generator
$h$ of $x^0$-translations, that is defined \cite{ap} as the
operator $h\psi:=i\hbar\dot\psi$ acting in the space ${\cal V}$ of
Klein-Gordon fields, squares to $\hbar^2\hat D$. Hence, as noted
in Ref.~\cite{p57}, $\hbar^{-1}D^{-1/2}h$ is nothing but the
charge-conjugation operator ${\cal C}$. This in turn means that
    \be
    iD^{-1/2}\dot\psi={\cal C}\psi=:\psi_c.
    \label{psi-c}
    \ee
Clearly $\psi_c$ is also a scalar field, and consequently
$D^{-1/2}\dot\psi$ is Lorentz invariant;
    \be
    \hat {D'}^{-1/2}\dot\psi'(x')=\hat D^{-1/2}\dot\psi(x).
    \label{scalar-2}
    \ee

Next, we use (\ref{boost}) and (\ref{scalar}) to deduce
    \bea
    \dot\psi'(x')&=&\dot\psi(x)+\vec\beta
    \cdot\vec\nabla\psi(x),
    \label{q1}\\
    \hat{D'}^\alpha&=&\hat D^\alpha-2\alpha\vec
    \beta\cdot\vec\nabla\hat D^{\alpha-1}\partial_0~~~~~~~~~~~~~
    \forall \alpha\in\R.
    \label{q2}
    \eea
In view of Eqs.~(\ref{scalar-2}) -- (\ref{q2}), we then have
    \bea
    \psi'(x')^*\hat{D'}^{1/2}\psi'(x')&=&
    \psi(x)^*\hat D^{1/2}\psi(x)-\psi(x)^*\vec\beta\cdot\vec\nabla
    \hat D^{-1/2}\dot\psi(x),
    \label{q3}\\
    \dot\psi'(x')^*\hat{D'}^{-1/2}\dot\psi'(x')&=&
    \dot\psi(x)^*\hat D^{-1/2}\dot\psi(x)+[\vec\beta\cdot\vec\nabla
    \psi(x)^*]\hat D^{-1/2}\dot\psi(x).
    \label{q4}
    \eea
Now, we substitute (\ref{scalar}), (\ref{q1}), (\ref{q3}), and
(\ref{q4}) in (\ref{j-zero-prime}) and make use of (\ref{boost-j})
to obtain
    \be
    \vec J_a(x)=\frac{\kappa}{2\cum}\left\{\psi(x)^*\vec\nabla
    \hat D^{-1/2}\dot\psi(x)-[\vec\nabla
    \psi(x)^*]\hat D^{-1/2}\dot\psi(x)-
    ia\left[\psi(x)^*\vec\nabla\psi(x)-[\vec\nabla\psi(x)^*]
    \psi(x)\right]\right\}.
    \label{j}
    \ee
This relation suggests
    \be
    J_a^\mu(x)=\frac{\kappa}{2\cum}\left\{\psi(x)^*
    \stackrel{\leftrightarrow}{\partial^\mu}\hat
    D^{-1/2}\dot\psi(x)-ia
    \psi(x)^*\stackrel{\leftrightarrow}{\partial^\mu}
    \psi(x)\right\}.
    \label{J}
    \ee
It is not difficult to check (using the Klein-Gordon equation)
that the expression for $J_a^0$ obtained using this equation
agrees with the one given in (\ref{j-zero}).

We can use (\ref{psi-c}) to further simplify (\ref{J}). This
yields
    \be
    J_a^\mu(x)=-\frac{i\kappa}{2\cum}\left[\psi(x)^*
    \stackrel{\leftrightarrow}{\partial^\mu}
    \tilde\psi_a(x)\right],
    \label{J-2}
    \ee
where
    \be
    \tilde\psi_a:=\psi_c+a\psi.
    \label{psi-tilde}
    \ee
Equation~(\ref{J-2}) is the main result of this
article.\footnote{Note that the results of Ref.~\cite{holland}
pertaining the uniqueness of the Klein-Gordon current density do
not rule out the existence of the current density (\ref{J}),
because these results are obtained under the assumption that the
current involves only the field and its first derivatives. The
appearance of $\hat D^{-1/2}$ in (\ref{J}) (alternatively $\psi_a$
in (\ref{J-2})) is a clear indication that this assumption is
violated.}

The current density $J_a^\mu$ constructed above has the following
remarkable properties.
    \begin{enumerate}
    \item The expression~(\ref{J-2}) for $J_a^\mu$ is manifestly
    covariant; since $\psi$ and $\tilde\psi_a$ are scalar fields,
    $J_a^\mu$ is indeed a four-vector field.
    \item Using the fact that both $\psi$ and $\tilde\psi_a$
    satisfy the Klein-Gordon equation (\ref{kg}), one can show (by
    a direct calculation) that the following continuity
    equation holds.
        \be
        \partial_\mu J_a^\mu=0.
        \label{conti-2}
        \ee
    Hence $J_a^\mu$ is a conserved current density.
    \item As we show in Appendix~B, in the nonrelativistic limit
    as $c\to\infty$, $J_a^\mu$ tends to the Schr\"odinger's
    probability current density for a free particle. Specifically,
    setting $\kappa=1/(1+a)$, we find
        \bea
        \lim_{c\to\infty} J_a^0(x^0,\vec x)=\varrho(x^0,\vec x),
        \label{nonrel-1}\\
        \lim_{c\to\infty} \vec J_a(x^0,\vec x)=\frac{1}{c}\;
        \vec j(x^0,\vec x),
        \label{nonrel-2}
        \eea
    where $\varrho$ and $\vec j$ are respectively the
    nonrelativistic scalar and current probability densities
    \cite{ryder}:
        \bea
        \varrho(x^0,\vec x)&:=&|\psi(x^0,\vec x)|^2,
        \label{class1}\\
        \vec j(x^0,\vec x)&:=&-\frac{i\hbar}{2m}
        \left[\psi(x^0,\vec x)^*\vec\nabla
        \psi(x^0,\vec x)-\psi(x^0,\vec x)\vec\nabla
        \psi(x^0,\vec x)^*\right].
        \label{class2}
        \eea
    \item Although $J_a^0(x)$ has been constructed out of a
    positive-definite inner product, namely~(\ref{gen-i}), it is
    in general not even real. This can be easily checked by
    computing $J_a^0(x)$ for a linear combination of two plane wave
    solutions of the Klein-Gordon equation with different and
    oppositely signed energies. Similarly $J_a^\mu$ is
    complex-valued. The real and imaginary parts of $J_a^\mu$ are by
    construction real-valued conserved 4-vector current densities.
    They are further studied in Appendix~C.
    \end{enumerate}

We can use the relation~(\ref{J}) for the current density
$J_a^\mu$ and Eq.~(\ref{gen-i}) to yield a manifestly covariant
expression for the most general positive-definite and
Lorentz-invariant inner product on the space of solutions of the
Klein-Gordon equation~(\ref{kg}), namely
    \be
    (\psi_1,\psi_2)_a=-\frac{i\kappa}{2\cum}
    \int_{\sigma} d\sigma(x)~ n_\mu(x)
    \left\{\psi_1(x)^*
    \stackrel{\leftrightarrow}{\partial^\mu}{\cal C}\psi_2(x)+a
    \psi_1(x)^*\stackrel{\leftrightarrow}{\partial^\mu}
    \psi_2(x)\right\},
    \label{man-cov}
    \ee
where $\sigma$ is an arbitrary spacelike (Cauchy) hypersurface of
the Minkowski space with volume element $d\sigma$ and unit
(future) timelike normal four-vector $n^\mu$. Note that in
deriving~(\ref{man-cov}) we have also made an implicit use of the
polarization principle \cite{kato}, namely that any inner product
is uniquely determined by the corresponding norm.

\section{Probability Current Density for Localization of
Klein-Gordon Fields in Space}

In nonrelativistic quantum mechanics, the interpretation of
$|\psi(\vec x;t)|^2$ as the probability density for the
localization of a particle in (configuration) space relies on the
following basic premises.
    \begin{itemize}
    \item[1.] The state of the particle is described by an
element $|\psi(t)\kt$ of the Hilbert space $L^2(\R^3)$.
    \item[2.] There is a Hermitian operator $\vec{\rm x}$
representing the position observable whose eigenvectors $|\vec
x\kt$ form a basis.
    \item[3.] The position wave function $\psi(\vec x;t)$ uniquely
determines the state vector $|\psi(t)\kt$ and consequently the
corresponding state, because $\psi(\vec x;t)$ are the coefficients
of the expansion of $|\psi(t)\kt$ in the position basis:
        \be
        |\psi(t)\kt=\int_{\R^3}d^3\vec x~\psi(\vec x;t)|\vec x\kt.
        \label{expand}
        \ee
    \item[4.] The probability of localization of the particle in a
region $V\subseteq\R^3$ at time $t\in\R$ is given by
        \be
        P_V(t)=\int_{V}d^3\vec x~\|\Lambda_{\vec x}|\psi(t)\kt\|^2,
        \label{project}
        \ee
where $\Lambda_x:=|\vec x\kt\br\vec x|$ is the projection operator
onto $|\vec x\kt$, $\|\cdot\|^2:=\br\cdot|\cdot\kt$, and the state
vector $|\psi(t)\kt$ is supposed to be normalized,
$\||\psi(t)\kt\|=1$. It is because of the orthonormality of the
position eigenvectors, i.e., $\br\vec x|\vec x'\kt=\delta^3(\vec
x-\vec x')$, that we can write (\ref{project}) in the form
        \be
        P_V(t)=\int_{V}d^3\vec x~|\psi(\vec x;t)|^2.
        \label{project2}
        \ee
    \end{itemize}

It is our belief that the same ingredients are necessary for
defining the probability density for the localization of the
Klein-Gordon fields in space, i.e., one must first define a
genuine Hilbert space and a position operator $\vec X$ for the
Klein-Gordon fields and then use the position eigenvectors to
define a position wave function associated with each Klein-Gordon
field. Refs.~\cite{ap,p57} give a thorough discussion of how one
can construct the Hilbert space, a position operator, and the
corresponding position wave functions for the Klein-Gordon and
similar fields. For completeness, here we include a brief summary
of this construction, elaborate on its consequences, and present
its application in our attempt to determine the probability
density for the localization of a Klein-Gordon field in space.

\subsection{The Hilbert Space}

Endowing the vector space ${\cal V}$ of Eq.~(\ref{v}) with the
inner product (\ref{gen-i}) and performing the Cauchy completion
of the resulting inner product space yield a separable Hilbert
space ${\cal H}_a$ for each choice of the parameter $a\in(-1,1)$.
However as discussed in great detail in \cite{ap}, the choice of
$a$ is physically irrelevant, because different choices yield
unitarily equivalent Hilbert spaces ${\cal H}_a$. In particular,
for all $a\in(-1,1)$, there is a unitary
transformation\footnote{The unitarity of $U_a$ means that for all
$\psi_1,\psi_2\in{\cal H}_a$, $\br
U_a\psi_1,U_a\psi_2\kt=(\psi_1,\psi_2)_a$ where
$\br\cdot,\cdot\kt$ stands for the inner product of
$L^2(\R^3)\oplus L^2(\R^3)$, \cite{reed-simon}.}
    \[U_a:{\cal H}_a\to L^2(\R^3)\oplus L^2(\R^3).\]
In fact, we can obtain the explicit form of $U_a$ rather easily.
In view of the general results of \cite{ap}, for all $\psi\in{\cal
H}_a$,
    \bea
    U_a\psi &:=&\frac{1}{2}\sqrt{\frac{\kappa}{\cum}}
            \left(\begin{array}{c}
            \sqrt{1+a}~[D^{1/4}\psi(x^0_0)+iD^{-1/4}\dot\psi(x^0_0)]\\
            \sqrt{1-a}~[D^{1/4}\psi(x^0_0)-iD^{-1/4}\dot\psi(x^0_0)]
            \end{array}\right)\nn\\
        &=&\frac{1}{2}\sqrt{\frac{\kappa}{\cum}}\,D^{1/4}
            \left(\begin{array}{c}
            \sqrt{1+a}~[\psi(x^0_0)+\psi_c(x^0_0)]\\
            \sqrt{1-a}~[\psi(x^0_0)-\psi_c(x^0_0)]
            \end{array}\right),
        \label{U-sub-a}
    \eea
where $x^0_0\in\R$ is a fixed initial value for $x^0$.

As far as the physical properties of the system are concerned we
can confine our attention to the simplest choice for $a$, namely
$a=0$. In this way we obtain the Hilbert space ${\cal H}:={\cal
H}_0$ that is mapped to $L^2(\R^3)\oplus L^2(\R^3)$ via the
unitary operator $U:=U_0$. Setting $a=0$ in (\ref{U-sub-a}) we
find \cite{p57}
    \be
    U\psi := \frac{1}{2}\sqrt{\frac{\kappa}{\cum}}
            \left(\begin{array}{c}
            D^{1/4}\psi(x^0_0)+iD^{-1/4}\dot\psi(x^0_0)\\
            D^{1/4}\psi(x^0_0)-iD^{-1/4}\dot\psi(x^0_0)
            \end{array}\right)
        =\frac{1}{2}\sqrt{\frac{\kappa}{\cum}}\,D^{1/4}
            \left(\begin{array}{c}
            \psi(x^0_0)+\psi_c(x^0_0)\\
            \psi(x^0_0)-\psi_c(x^0_0)
            \end{array}\right).
        \label{U}
    \ee
We can also calculate the inverse of $U$. The result is \cite{p57}
    \be
    \left[U^{-1} \xi\right](x^0)
    =\sqrt{\frac{\cum}{\kappa}}\;D^{-1/4}
       \left[e^{-i(x^0-x^0_0)D^{1/2}} \xi_1
       +e^{i(x^0-x^0_0)D^{1/2}} \xi_2 \right],
    \label{U-inv}
    \ee
where $\xi=\mbox{\tiny $\left(\begin{array}{c} \xi_1\\
\xi_2\end{array}\right)$}\in L^2(\R^3)\oplus L^2(\R^3)$ and
$x^0\in\R$ are arbitrary.

It is important to note that $U_a$ (and in particular $U$) depend
on the choice of $x^0_0$. Therefore they fail to be unique.

\subsection{Position and Momentum Operators}

Let $\vec{\rm x}$ and $\vec{\rm p}$ be the usual position and
momentum operators acting in $L^2(\R^3)$, respectively, $|\vec
x\kt$ be the position eigenvectors satisfying
    \be
    \vec{\rm x}|\vec x\kt=\vec x|\vec x\kt,~~~~~~
    \br\vec x|\vec x'\kt=\delta^3(\vec x-\vec x'),~~~~~
    \int_{\R^3}d^3\vec x~|\vec x\kt\br\vec x|=1,
    \label{x-prop}
    \ee
$\sigma_i$ with $i\in\{1,2,3\}$ be the Pauli matrices
    \be
    \sigma_1=\left(\begin{array}{cc}
    0 & 1 \\
    1 & 0\end{array}\right),~~~~~~~
    \sigma_2=\left(\begin{array}{cc}
    0 & -i \\
    i & 0\end{array}\right),~~~~~~~
    \sigma_3=\left(\begin{array}{cc}
    1 & 0 \\
    0 & -1\end{array}\right),
    \label{pauli}
    \end{equation}
and $\sigma_0$ be the $2\times 2$ identity matrix. Then any
observable acting in $L^2(\R^3)\oplus L^2(\R^3)$ is of the form
$O=\sum_{\mu=0}^3 O_\mu\otimes\sigma_\mu$ where $O_\mu$ are
Hermitian operators acting in $L^2(\R^3)$. This in turn implies
that the general form of the observables (Hermitian operators)
acting in the Hilbert space ${\cal H}$ is given by $U^{-1}OU$, for
$U:{\cal H}\to L^2(\R^3)\oplus L^2(\R^3)$ is a unitary operator.
In particular, as proposed in \cite{p57}, we may identify the
operators $\vec X,\vec P:{\cal H}\to{\cal H}$, defined by
    \be
    \vec X:= U^{-1}(\vec{\rm x}\otimes\sigma_0)U,~~~~~~~
    \vec P:= U^{-1}(\vec{\rm p}\otimes\sigma_0)U,
    \label{xp}
    \ee
with position and momentum operators for the Klein-Gordon fields,
respectively. It turns out that $(\vec P\psi)(x^0)=\vec{\rm
p}\psi(x^0)$ but that $\vec X\psi$ has a more complicated
expression. It is determined by the initial conditions \cite{p57}
    \[ (\vec X\psi)(x_0^0)=\vec{\cal X}\psi(x_0^0),~~~~~~~~
    \partial_0(\vec X\psi)(x_0^0)=\vec{\cal X}^\dagger
    \dot\psi(x_0^0),\]
where
    \[ \vec{\cal X}:=\vec{\rm x}+\frac{i\hbar\vec{\rm
    p}}{2(\vec{\rm p}^2+m^2)}\]
is the Newton-Wigner position operator~\cite{newton-wigner}.

\subsection{Localized States and Position Wave Functions}

Clearly the operators $\vec{\rm x}\otimes\sigma_0$ and
$1\otimes\sigma_3$ from a maximal commuting set of observables
acting in $L^2(\R^3)\oplus L^2(\R^3)$. Hence their common
eigenvectors
    \be
    \xi^{(\epsilon,\vec x)}:=|\vec x\kt\otimes e_\epsilon
    \label{basis-xi}
    \ee
with $e_+:=\mbox{\tiny $\left(\begin{array}{c} 1\\
0\end{array}\right)$}$ and $e_-:=\mbox{\tiny $\left(\begin{array}{c} 0\\
1\end{array}\right)$}$, form a complete orthonormal basis of
$L^2(\R^3)\oplus L^2(\R^3)$. This together with the fact that $U$
is a unitary transformation imply that
    \be
    \psi^{(\epsilon,\vec x)}:=U^{-1}\xi^{(\epsilon,\vec x)}
    \label{basis-psi}
    \ee
form a complete orthonormal basis of ${\cal H}$, i.e.,
    \be
    (\psi^{(\epsilon,\vec x)},\psi^{(\epsilon',\vec x')})_0
    =\delta_{\epsilon,\epsilon'}\delta^3(\vec x-\vec x'),~~~~~~
    \sum_{\epsilon=\pm 1}\int_{\R^3}d^3\vec x~
    |\psi^{(\epsilon,\vec x)})(\psi^{(\epsilon,\vec x)}|=I,
    \label{com-ort}
    \ee
where for all $\psi\in{\cal H}$, $|\psi)(\psi|$ is the projection
operator defined by $|\psi)(\psi|\phi:=(\psi,\phi)_0\psi$ and $I$
is the identity map acting in ${\cal H}$. Furthermore, we have for
both $\epsilon=\pm 1$
    \be
    \vec X \psi^{(\epsilon,\vec x)}=\vec x \psi^{(\epsilon,\vec
    x)}.
    \label{X-psi}
    \ee
It is also not difficult to see \cite{p57} that the
charge-conjugation transformation is given by ${\cal
C}=U^{-1}(1\otimes\sigma_3)U$. Hence,
    \be
    {\cal C} \psi^{(\epsilon,\vec x)}=\epsilon \psi^{(\epsilon,\vec
    x)}.
    \label{C-psi}
    \ee

In view of Eqs.~(\ref{com-ort}) -- (\ref{C-psi}), the state
vectors $\psi^{(\epsilon,\vec x)}$ represent spatially localized
Klein-Gordon fields with definite charge-parity $\epsilon$. They
can be employed to associate each Klein-Gordon field $\psi\in{\cal
H}$ with a unique position wave function, namely
    \be
    f(\epsilon,\vec x):=(\psi^{(\epsilon,\vec x)},\psi)_0.
    \label{f}
    \ee
As shown in \cite{p57}, one can use these wave functions to
represent all the physical quantities associated with the
Klein-Gordon fields. In particular, the transition amplitudes
between two states (inner product of two state vectors) take the
simple form
    \be
    (\psi_1,\psi_2)_0=\sum_{\epsilon=\pm 1}\int_{\R^3}d^3\vec x~
    f_1(\epsilon,\vec x)^*f_2(\epsilon,\vec x),
    \label{inn-wf}
    \ee
where $\psi_1,\psi_2\in{\cal H}$ and $f_1,f_2$ are the
corresponding wave functions.

As suggested by (\ref{inn-wf}), the wave functions $f(\pm,\vec x)$
belong to $L^2(\R^3)$. Moreover due to the implicit dependence of
$\psi^{(\epsilon,\vec x)}$ on $x_0^0$ appearing in the expression
for $U$, $f(\pm,\vec x)$ depend on $x_0^0$. This dependence
becomes explicit once we express $f(\epsilon,\vec x)$ in terms of
$\psi$ directly. In order to see this, we first substitute
(\ref{U-inv}) and (\ref{basis-xi}) in (\ref{basis-psi}) to obtain
    \be
    \psi^{(\epsilon,\vec x)}(x^0)=
    \sqrt{\frac{\cum}{\kappa}}~
     D^{-1/4} e^{-i\epsilon(x^0-x^0_0)D^{1/2}}|\vec x\kt.
    \label{b-1}
    \ee
We then use this equation and (\ref{gen-i}) to compute the
right-hand side of (\ref{f}). This yields
    \be
    f(\epsilon,\vec x)=\sqrt{\frac{\kappa}{\cum}}~ \hat D^{1/4}
    e^{i\epsilon(x^0-x^0_0)\hat D^{1/2}}
    \psi_\epsilon(x^0,\vec x),
    \label{f2}
    \ee
where
    \be
    \psi_\epsilon:=\frac{1}{2}(1+\epsilon{\cal C})\psi=
    \frac{1}{2}(\psi+\epsilon\psi_c)
    \label{psi-pm}
    \ee
is the definite-charge (definite-energy) component of $\psi$ with
charge-parity $\epsilon$. Note however that $\psi_\epsilon$
satisfies the Foldy equation \cite{foldy}
    \be
    i\partial_0\psi_\epsilon(x^0,\vec x)=\epsilon\hat D^{1/2}
    \psi_\epsilon(x^0,\vec x).
    \label{foldy}
    \ee
This in turn implies
    \[e^{i\epsilon(x^0-x^0_0)\hat D^{1/2}}
    \psi_\epsilon(x^0,\vec x)=\psi_\epsilon(x_0^0,\vec x).\]
Hence (\ref{f2}) takes the simple form
    \be
    f(\epsilon,\vec x)=\sqrt{\frac{\kappa}{\cum}}~ \hat D^{1/4}
    \psi_\epsilon(x_0^0,\vec x).
    \label{f3}
    \ee
As seen from this equation the wave functions $f(\epsilon,\vec x)$
depend on $x_0^0$.

It is also interesting to note that one can use (\ref{b-1}) to
compute
    \bea
    \psi^{(\epsilon,\vec y)}(x^0,\vec x)&:=&
    \br\vec x|\psi^{(\epsilon,\vec y)}(x^0)\kt\nn\\
    &=&
    \sqrt{\frac{\cum}{\kappa}}~\frac{1}{2\pi^2|\vec x-\vec y|}
    \int_0^\infty dk\left\{
    \frac{k\,\sin(|\vec x-\vec y|k)\;
    \exp\left[-i\epsilon(x^0-x_0^0)\sqrt{k^2+\cum^2}\right]}{
    (k^2+\cum^2)^{1/4}}\right\}.
    \nn
    \eea
For $x^0=x_0^0$, the integral on the right-hand side of this
equation can be expressed in terms of the Bessel K-function
$K_\frac{5}{4}$. The result is, for both $\epsilon=-1$ and $1$,
    \be
    \psi^{(\epsilon,\vec y)}(x_0^0,\vec x)=
    \sqrt{\frac{\cum}{\kappa}}~\left[2^\frac{3}{4}\pi^\frac{3}{2}
    \Gamma(\mbox{\footnotesize$\frac{1}{4}$})\right]^{-1}\left(\frac{\cum}{|\vec x-\vec
    y|}\right)^\frac{5}{4} K_\frac{5}{4}(\cum|\vec x-\vec y|),
    \label{exn2}
    \ee
where $\Gamma$ stands for the Gamma function.

Equation~(\ref{exn2}) provides an explicit demonstration of the
curious fact that $\psi^{(+,\vec x)}$ are indeed identical with
the Newton-Wigner localized states \cite{newton-wigner} and that
$\psi^{(-,\vec x)}$ are the negative-energy analogs of the latter.
It is remarkable that we have obtained these localized states
without pursuing the axiomatic approach of
Ref.~\cite{newton-wigner}. A perhaps more important observation is
that actually one does not need to use the rather complicated
expression (\ref{exn2}) in calculating physical quantities
\cite{ap,p57}. One can instead employ the corresponding wave
functions which are simply delta functions: The wave function
$f_{(\epsilon,\vec x)}(\epsilon',\vec x')$ for
$\psi^{(\epsilon,\vec x)}(x_0^0)$ has the form
$\delta_{\epsilon,\epsilon'}\delta(\vec x-\vec x')$.

\subsection{Probability Density for Spatial Localization of a Field}

Having obtained the expression for the position operator $\vec X$
and position wave functions $f(\epsilon,\vec x)$, we may proceed
as in nonrelativistic quantum mechanics and identify the
probability of the localization of a Klein-Gordon field $\psi$ in
a region $V\subseteq\R^3$, at time $t_0=x_0^0/c$, with
    \be
    P_V=\int_V d^3\vec x~\|\Pi_{\vec x}\psi\|_0^2,
    \label{project-kg}
    \ee
where $\Pi_{\vec x}$ is the projection operator onto the
eigenspace of $\vec X$ with eigenvalue $\vec x$, i.e.,
    \[\Pi_{\vec x}=\sum_{\epsilon=\pm 1}
    |\psi^{(\epsilon,\vec x)})(\psi^{(\epsilon,\vec x)}|,\]
$\|\cdot\|_0^2:=(\cdot,\cdot)_0$ is the square of the norm of
${\cal H}$, and we assume $\|\psi\|_0=1$. Substituting this
relation in (\ref{project-kg}) and making use of (\ref{com-ort})
and (\ref{f}), we have
    \[P_V=\sum_{\epsilon=\pm 1}\int_V d^3\vec x~|f(\epsilon,\vec
    x)|^2.\]
Therefore, the probability density is given by
    \be
    \rho(x_0^0,\vec x)=\sum_{\epsilon=\pm 1}
    |f(\epsilon,\vec x)|^2=\frac{\kappa}{2\cum}
    \left\{ |\hat D^{1/4}\psi(x_0^0,\vec x)|^2+
    |\hat D^{-1/4}\dot\psi(x_0^0,\vec x)|^2\right\}.
    \label{rho=}
    \ee
To establish the second equality in (\ref{rho=}), we have made use
of (\ref{f3}), (\ref{psi-pm}), and (\ref{psi-c}). For a position
measurement to be made at time $t=x^0/c$, we have the probability
density
    \be
    \rho(x^0,\vec x)=\frac{\kappa}{2\cum}\left\{
    |\hat D^{1/4}\psi(x^0,\vec x)|^2+
    |\hat D^{-1/4}\dot\psi(x^0,\vec x)|^2\right\}.
    \label{rho}
    \ee
We can use (\ref{psi-c}) to express $\rho$ in the following
slightly more symmetrical form.
    \be
    \rho(x)=\frac{\kappa}{2\cum}\left\{
    |\hat D^{1/4}\psi(x)|^2+|\hat D^{1/4}\psi_c(x)|^2\right\}.
    \label{rho-c}
    \ee

Although the above discussion is based on a particular choice for
the parameter $a$, namely $a=0$, it is generally valid. To see
this, suppose we choose to work with the inner product
(\ref{gen-i}) and hence the Hilbert space ${\cal H}_a$ for some
$a\neq 0$. Then we have a different position operator: $\vec
X_a:=U_a^{-1}(\vec{\rm x}\otimes\sigma_0)U_a={\cal U}_a\vec X{\cal
U}_a^{-1}$ where
    \be
    {\cal U}_a:=U_a^{-1}U
    \label{cur-u-a}
    \ee
is a unitary operator mapping ${\cal H}$ onto ${\cal H}_a$. The
eigenvectors $\psi_a^{(\epsilon,\vec x)}$ of $\vec X_a$ are
clearly related to $\psi^{(\epsilon,\vec x)}$ by
$\psi_a^{(\epsilon,\vec x)}={\cal U}_a \psi^{(\epsilon,\vec x)}$.
Now, given $\psi_a\in{\cal H}_a$, we define $\psi:={\cal
U}_a^{-1}\psi_a$ and check that the position wave function for
$\psi_a$ is given by
        \be
        f_a(\epsilon,\vec x):=(\psi^{(\epsilon,\vec
        x)}_a,\psi_a)_a=({\cal U}_a\psi^{(\epsilon,\vec
        x)},{\cal U}_a\psi)_a=(\psi^{(\epsilon,\vec
        x)},\psi)=f(\epsilon,\vec x).
        \label{f=f}
        \ee
Here we made use of the fact that ${\cal U}_a:{\cal H}\to{\cal
H}_a$ is a unitary operator. As seen from (\ref{f=f}), the
position wave functions for $\psi$ and $\psi_a$ coincide. As a
result so do the corresponding probability densities.

If we are to compute the probability density $\rho_a$ of the
spatial localization of a Klein-Gordon field $\psi$ with the
position operator being identified with $\vec X_a$ for $a\neq 0$,
we have, for a measurement made at $t_0=x_0^0/c$,
    \be
    \rho_a(x_0^0,\vec x)=\frac{\kappa}{2\cum}\left\{
    |\hat D^{1/4}\psi'_a(x_0^0,\vec x)|^2+
    |\hat D^{-1/4}\dot\psi'_a(x_0^0,\vec x)|^2\right\}
    \label{rho-a}
    \ee
where $\psi'_a:={\cal U}_a^{-1}\psi$. We can compute the latter
using (\ref{U-sub-a}), (\ref{U}), and (\ref{cur-u-a}). This leads
to
    \be
    \psi'_a(x^0_0)=\alpha_+\psi(x_0^0)+i\alpha_-\hat D^{-1/2}
    \dot\psi(x_0^0),~~~~~~~~~
    \dot\psi'_a(x^0_0)=-i\alpha_-\hat D^{1/2}\psi(x_0^0)+
    \alpha_+\dot\psi(x_0^0),
    \label{psi-psi}
    \ee
where
    \be
    \alpha_\pm:=\frac{1}{2}~\left(\sqrt{1+a}\pm\sqrt{1-a}\right).
    \label{alpha=}
    \ee
Now, substituting (\ref{psi-psi}) and (\ref{alpha=}) in
(\ref{rho-a}) and doing the necessary algebra, we find the
following remarkably simple result.
    \bea
    \rho_a(x_0^0,\vec x)&=&\frac{\kappa}{2\cum}\left\{
    |\hat D^{1/4}\psi(x_0^0,\vec x)|^2+
    |\hat D^{-1/4}\dot\psi(x_0^0,\vec x)|^2+\right.\nn\\
    &&\vspace{2cm}\left.ia\left[(\hat D^{1/4}\psi(x_0^0,\vec x))^*
    \hat D^{-1/4}\dot\psi(x_0^0,\vec x)-
    (\hat D^{1/4}\psi(x_0^0,\vec x))
    (\hat D^{-1/4}\dot\psi(x_0^0,\vec x))^*\right]\right\}\nn\\
    &=&\frac{\kappa}{2\cum}\left\{
    |\hat D^{1/4}\psi(x_0^0,\vec x)|^2+
    |\hat D^{1/4}\psi_c(x_0^0,\vec x)|^2+\right.\nn\\
    &&\vspace{2cm}\left.a\left[(\hat D^{1/4}\psi(x_0^0,\vec x))^*
    \hat D^{1/4}\psi_c(x_0^0,\vec x)+(\hat D^{1/4}\psi(x_0^0,\vec x))
    (\hat D^{1/4}\psi_c(x_0^0,\vec x))^*\right]\right\}.
    \label{rho-a-2}
    \eea
For a measurement made at $t=x^0/c$ we therefore have
    \bea
    \rho_a(x)&=&\frac{\kappa}{2\cum}\left\{
    |\hat D^{1/4}\psi(x)|^2+
    |\hat D^{-1/4}\dot\psi(x)|^2+
    \right.\nn\\&&\vspace{2cm}\left.
    ia\left[(\hat D^{1/4}\psi(x))^*
    \hat D^{-1/4}\dot\psi(x)-(\hat D^{1/4}\psi(x))
    (\hat D^{-1/4}\dot\psi(x))^*\right]\right\}\nn\\
    &=&\frac{\kappa}{2\cum}\left\{
    |\hat D^{1/4}\psi(x)|^2+
    |\hat D^{1/4}\psi_c(x)|^2+
    \right.\nn\\&&\vspace{2cm}\left.
    a\left[(\hat D^{1/4}\psi(x))^*
    \hat D^{1/4}\psi_c(x)+(\hat D^{1/4}\psi(x))
    (\hat D^{1/4}\psi_c(x))^*\right]\right\}.
    \label{rho-a-3}
    \eea

For a positive-energy Klein-Gordon field, (\ref{rho-c}) reduces to
a probability density originally introduced by Rosenstein and
Horwitz \cite{rh} by restricting the second-quantized scalar field
theory to its one-particle sector. This coincidence may be viewed
as a verification of the validity of our approach: The
first-quantized theory formulated by an explicit construction of
the Hilbert space and a position observable reproduces a result
obtained from the second-quantized theory.

We can use the method discussed in Section~2 to also define a
current density ${\cal J}_a^\mu$ such that ${\cal J}_a^0=\rho_a$.
As we show in Appendix~D, this yields
    \bea
   {\cal J}_a^\mu(x)&=&\frac{\kappa}{2\cum}\Im\left\{
   (\hat D^{1/4}\psi(x))^* \partial^\mu \hat D^{-1/4}\psi_c(x) -
   (\hat D^{1/4}\psi_c(x)) \partial^\mu (\hat D^{-1/4}\psi(x))^* +
   \right.\nn\\
   &&\vspace{2cm}\left.a\left[
   (\hat D^{1/4}\psi(x))^* \partial^\mu \hat D^{-1/4}\psi(x) -
   (\hat D^{1/4}\psi_c(x)) \partial^\mu (\hat D^{-1/4}
   \psi_c(x))^*\right]\right\},
   \label{calj-3-n}
   \eea
where $\Im$ stands for the imaginary part of its argument.
Furthermore, the probability current density ${\cal J}_a^\mu$ has
the correct nonrelativistic limit: Setting $\kappa=1/(1+a)$ yields
    \be
    \lim_{c\to\infty} {\cal J}_a^0(x^0,\vec x)=\varrho
    (x^0,\vec x),~~~~~~~~~
    \lim_{c\to\infty} \vec{\cal J}_a(x^0,\vec x)=\frac{1}{c}\;
        \vec j(x^0,\vec x),
    \label{class3}
    \ee
where $\varrho$ and $j$ are the classical scalar and current
probability densities given by (\ref{class1}) and (\ref{class2}),
respectively.

If we restrict to the positive-energy Klein-Gordon fields and set
$a=0$, Eq.~(\ref{calj-3-n}) reduces to the probability current
density obtained by Rosenstein and Horwitz in \cite{rh}. As also
indicated by these authors, the resulting current density, namely
${\cal J}_0^\mu$, does not satisfy the continuity equation. Hence
it is not a conserved current. Furthermore, as we show in
Appendix~D, ${\cal J}_0^\mu$ is indeed not even a four-vector
field. The same lack of covariance and conservation applies to
${\cal J}_a^\mu$ for $a\neq 0$. The only advantage of ${\cal
J}_a^\mu$ over $J_a^\mu$ is that, unlike the latter which is
generally complex-valued, the former is manifestly real-valued and
positive-definite.

The non-conservation (respectively non-covariance) of the
probability current density ${\cal J}_a^\mu$ raises the
paradoxical possibility of the non-conservation (respectively
frame-dependence) of the total probability:
    \be
    {\cal P}_a:=\int_{\R^3}d^3\vec x~\rho_a(x^0,\vec x).
    \label{Prob-00}
    \ee
It turns out that indeed the latter is a frame-independent
conserved quantity, thanks to the covariance and conservation of
the current density $J_a^\mu$ and the identity
    \be
    \int_{\R^3}d^3\vec x~\rho_a(x^0,\vec x)=
    \int_{\R^3}d^3\vec x~J_a^0(x^0,\vec x),
    \label{Prob-01}
    \ee
which follows from (\ref{j-zero}), (\ref{rho-a-3}) and the fact
that $D^{\pm 1/4}$ is a self-adjoint operator acting in $L^2(\R)$.
In a sense, $\rho_a(x)$ and $J_a^0(x)$ differ only by a ``boundary
term''.

Combining (\ref{Prob-00}) and (\ref{Prob-01}), we have
    \be
    {\cal P}_a=\int_{\R^3}d^3\vec x~J_a^0(x^0,\vec x).
    \label{rho=j}
    \ee
This relation implies that although the probability density
$\rho_a$ is not the zero-component of a conserved four-vector
current density, its integral over the whole space that yields the
total probability (\ref{Prob-00}) is nevertheless conserved.
Furthermore, this global conservation law stems from a local
conservation law, i.e., a continuity equation for a four-vector
current density namely $J_a^\mu$.

\section{Gauge Symmetry Associated with the Conservation of the
Total Probability}

The fact that the conservation of the total probability ${\cal
P}_a$ has its root in the local conservation of the covariant
current $J^\mu_a$ suggests, by virtue of the N\"other's theorem,
the presence of an underlying gauge symmetry. In order to
determine the nature of this symmetry, we make use of the
well-known fact that the conserved charge associated with any
conserved current is the generator of the infinitesimal gauge
transformations \cite{itzykson-zuber}. The specific form of the
latter is most conveniently obtained in the Hamiltonian
formulation.

The Lagrangian $L$ for a free Klein-Gordon field $\psi$ and the
corresponding canonical momenta $\pi(\vec x)$, $\bar\pi(\vec x)$
associated with $\psi(\vec x):=\psi(x^0,\vec x)$ and $\psi^*(\vec
x):=\psi^*(x^0,\vec x)$ are respectively given by \cite{weinberg}:
    \bea
    L&:=&-\frac{\lambda}{2}\int_{\R^3}d^3\vec x\left\{
    \partial_\mu \psi(\vec x)^*\partial^\mu\psi(\vec x)+
    \cum^2\psi(\vec x)^*\psi(\vec x)\right\},
    \label{Lag}\\
    \pi(\vec x)&:=&\frac{\delta L}{\delta\dot\psi(\vec x)}=
    \frac{\lambda}{2}~\dot\psi^*(\vec x),~~~~~~
    \bar\pi(\vec x):=\frac{\delta L}{\delta\dot\psi^*(\vec x)}
    =\frac{\lambda}{2}~\dot\psi(\vec x)=\pi^*(\vec x),
    \label{mom-star}
    \eea
where $\lambda:=\hbar c/\cum=\hbar^2/m$ and we have suppressed the
$x^0$-dependence of the fields for simplicity. In terms of the
canonical phase space variables $(\psi,\pi)$ and $(\psi^*,\pi^*)$,
the conserved charge for $J^\mu_a$, namely the total
probability~(\ref{rho=j}), takes the form
    \bea
    {\cal P}_a&=&\frac{\kappa}{2\cum}\int_{\R^3}d^3\vec x\:\left\{
    \psi(\vec x)^* \hat D^{1/2}\psi(\vec x)+
    4\lambda^{-2}\pi(\vec x)\hat D^{-1/2}
    \pi^*(\vec x)+\right.\nn\\
    &&\hspace{3cm}\left. 2i\lambda^{-1}a[\psi(\vec x)^*
    \pi(\vec x)^*-\psi(\vec x)\pi(\vec x)]\right\},
    \label{prob=}
    \eea
where we have made use of (\ref{j-zero}) and (\ref{mom-star}).

We can obtain the infinitesimal symmetry transformation,
    \be
    \psi\to\psi+\delta\psi,
    \label{trans}
    \ee
generated by ${\cal P}_a$ using
    \be
    \delta\psi(\vec x)=\left\{\psi(\vec x),{\cal
    P}_a\right\}\;\delta\xi,
    \label{var}
    \ee
where $\left\{\cdot,\cdot\right\}$ is the Poisson bracket:
    \be
    \left\{ {\cal A},{\cal B}\right\}:=
    \int_{\R^3}d^3\vec x\:\left[
    \frac{\delta{\cal A}}{\delta\psi(\vec x)}
    \frac{\delta{\cal B}}{\delta\pi(\vec x)}
    -\frac{\delta{\cal B}}{\delta\psi(\vec x)}
    \frac{\delta{\cal A}}{\delta\pi(\vec x)}
    +\frac{\delta{\cal A}}{\delta\psi^*(\vec x)}
    \frac{\delta{\cal B}}{\delta\pi^*(\vec x)}
    -\frac{\delta{\cal B}}{\delta\psi^*(\vec x)}
    \frac{\delta{\cal A}}{\delta\pi^*(\vec x)}\right],
    \label{poisson}
    \ee
${\cal A},{\cal B}$ are observables, and $\delta\xi$ is an
infinitesimal real parameter. In view of (\ref{mom-star}) --
(\ref{poisson}), we have
    \[\delta\psi(\vec x)=\frac{\delta{\cal P}_a}{\delta\pi(\vec x)}
    \;\delta\xi
    =\frac{\kappa}{\cum\lambda}\left[
    \hat D^{-1/2}\dot\psi(\vec x)-ia\psi(\vec x)\right]\delta\xi.\]
We may employ (\ref{psi-c}) to further simplify this expression.
The result is
    \be
    \delta\psi(\vec x)=-i\delta\theta~({\cal C}+a)\psi(\vec x),
    \label{gauge-1}
    \ee
where $\delta\theta:=\kappa\,\delta\xi/(\cum\lambda)=
\kappa\,\delta\xi/(\hbar c)$.

According to (\ref{gauge-1}), the symmetry transformations
(\ref{trans}) are generated by the operator ${\cal C}+a$. One can
easily exponentiate the latter to obtain the following expression
for the corresponding non-infinitesimal symmetry transformations.
    \be
    \psi\to e^{-i\theta({\cal C}+a)}\psi=
        e^{-ia\theta} e^{-i\theta{\cal C}}\psi=
        e^{-ia\theta}[\cos\theta-i\sin\theta~{\cal C}]\psi.
    \label{gauge-2}
    \ee
where $\theta\in\R$ is arbitrary and we have made use of ${\cal
C}^2=1$. In terms of the positive- and negative-energy components
$\psi_\pm$ of $\psi$. The expression~(\ref{gauge-2}) takes the
form
    \be
    \psi=\psi_++\psi_-\to e^{-i(a+1)\theta}\psi_++
    e^{-i(a-1)\theta}\psi_-=
    \sum_{\epsilon=\pm} e^{-i(a+\epsilon)\theta}\psi_\epsilon.
    \label{gauge-3}
    \ee

It is not difficult to see from (\ref{gauge-2}) and
(\ref{gauge-3}) that the gauge group\footnote{Here we identify the
gauge group with its connected component that includes the
identity and is obtained by exponentiating the generator ${\cal
C}+a$.} $G_a$ associated with these transformations is a
one-dimensional connected Abelian Lie group. Therefore, it is
isomorphic to either of $U(1)$ or $\R^+$, the latter being the
noncompact multiplicative group of positive real numbers,
\cite{brocker-dieck}.

We can construct a simple model for (faithful representation of)
the group $G_a$ using the two-component representation
$\psi=\mbox{\scriptsize$\left(\begin{array}{c}\psi_+\\ \psi_-
\end{array}\right)$}$. Then ${\cal C}$ is represented by the
diagonal Pauli matrix
$\sigma_3=\mbox{\scriptsize$\left(\begin{array}{cc}1 & 0 \\0 &
-1\end{array}\right)$}$ and a typical element of $G_a$ takes the
form
    \be
    g_a(\theta):=\left(\begin{array}{cc}
    e^{-i(a+1)\theta} & 0 \\
    0 & e^{-i(a-1)\theta}\end{array}\right).
    \label{element}
    \ee
This expression suggests that the gauge group $G_a$ is a subgroup
of $U(1)\times U(1)$. It is not difficult to show that $G_a$ is a
compact subgroup of this group and consequently isomorphic to the
group $U(1)$ if and only if the parameter $a$ is a rational
number. This in turn implies that for irrational $a$ the group
$G_a$ is isomorphic to $\R^+$.\footnote{In this case, although
$G_a$ is (isomorphic to) an abstract subgroup of $U(1)\times U(1)$
it fails to be a Lie subgroup of this group \cite{brocker-dieck}.}

We can easily construct a concrete example of these isomorphisms.
For a rational $a$, we have $a=m/n$ where $m$ and $n$ are
relatively prime integers with $n$ positive. In this case we let
$u_a:G_a\to U(1)$ be defined by $u_a(e^{-i\theta(a+{\cal
C})}):=e^{-i\theta/n}$. For an irrational $a$, we define
$v_a:G_a\to \R^+$ according to $v_a(e^{-i\theta(a+{\cal
C})}):=e^\theta $. Then it is an easy exercise to show that both
$u_a$ and $v_a$ are (Lie) group isomorphisms.

Clearly, the $G_a$ gauge symmetry associated with the conservation
of the total probability, alternatively the current density
$J_a^\mu$, is a global gauge symmetry. Similarly to the $U(1)$
gauge symmetry associated with the Klein-Gordon current, namely
the one responsible for the electric charge conservation, one may
consider allowing for $(x^0,\vec x)$-dependent gauge parameters:
$\theta=\theta(x)$, i.e., consider local $G_a$ gauge
transformations. One then expects that the imposition of this
local gauge symmetry should lead to a gauged Klein-Gordon equation
involving a gauge field that couples to the current $J^\mu_a$. The
naive minimal coupling prescription, however, fails because it
makes the generator ${\cal C}+a$ of $G_a$ generally $(x^0,\vec
x)$-dependent. In this respect the local $G_a$ gauge symmetry is
different from the usual local Yang-Mills-type gauge symmetries.

The group $U(1)\times U(1)$ that enters the above discussion of
the gauge group $G_a$ as an embedding group is also a group of
gauge transformations of the Klein-Gordon fields. It corresponds
to the global $G_a$ gauge transformations supplemented with the
global $U(1)$ gauge transformations associated with the
conservation of the electric charge (\ref{q}).

\section{Klein-Gordon Fields in a Background Electromagnetic Field}

A scalar field $\psi$ minimally coupled to a background
electromagnetic field $A_\mu$ satisfies
    \be
    \Big\{[-i\partial_\mu-q A_\mu(x)][-i\partial^\mu-q
    A^\mu(x)]+\cum^2\Big\}\psi(x)=0,
    \label{ckg}
    \ee
where $q:=e/(\hbar c)$, $e$ is the electric charge, and $A_\mu$ is
assumed to be real-valued.

Supposing that for all $x^0$, $\psi(x^0,\vec x)$ and
$\dot\psi(x^0,\vec x)$ are square-integrable functions, we can
easily write Eq.~(\ref{ckg}) as an ordinary differential equation
in $L^2(\R^3)$. The latter takes the form
    \be
    \ddot\psi(x^0)+2iq\:\varphi(x^0,\vec{\rm x})~\dot\psi(x^0)+
    {\cal D}\,\psi(x^0)=0,
    \label{ckg-ode}
    \ee
where $\varphi:=A^0$ is the scalar potential, ${\cal
D}:L^2(\R^3)\to L^2(\R^3)$ is the operator:
    \be
    ({\cal D}\phi)(\vec x):=\left\{-[\vec\nabla-iq\vec
    A(x^0,\vec x)]^2+iq\dot\varphi(x^0,\vec x)-q^2
    \varphi(x^0,\vec x)^2+\cum^2\right\}\phi(\vec x)~~~~~~~~~~~
    \forall\phi\in L^2(\R^3),
    \label{cud=}
    \ee
and $\vec A=(A^1,A^2,A^3)$ is the vector potential.

Equation~(\ref{ckg-ode}) takes the form of the Klein-Gordon
equation~(\ref{kg-0}) for a free scalar field provided that we
make the gauge transformation:
    \be
    \psi(x^0)\to \chi(x^0):=u(x^0,\vec{\rm x})\psi(x^0),~~~~~~~~~~~~
    u(x^0,\vec x):=\exp\left[iq\int_{x^0_0}^{x^0}d\tau~
    \varphi(\tau,\vec{x})\right],
    \label{gauge1}
    \ee
where $x_0^0\in\R$ is an arbitrary but fixed initial value of
$x^0$. Substituting (\ref{cud=}) and (\ref{gauge1}) in
(\ref{ckg-ode}) and doing the necessary algebra, we find
    \be
    \ddot\chi(x^0)+D_q\:\chi(x^0)=0,
    \label{kgt-gauge}
    \ee
where $D_q:L^2(\R^3)\to L^2(\R^3)$ is the operator
    \be
    (D_q\phi)(\vec x):=\hat D_q\phi(\vec x)~~~~~~~~~~~~
    \forall\phi\in L^2(\R^3),
    \label{Dq1}
    \ee
and
    \be
    \hat D_q:=u(x^0,\vec x)\left\{-[\vec\nabla-
    iq\vec A(x^0,\vec x)]^2+\cum^2\right\}u(x^0,\vec x)^{-1}.
    \label{Dq2}
    \ee

It is not difficult to observe that indeed $D_q$ is a
positive-definite operator acting in $L^2(\R^3)$. Therefore,
according to the terminology of Refs.~\cite{cqg,ap},
(\ref{kgt-gauge}) is an example of a Klein-Gordon-type field
equation. The main difference between the free Klein-Gordon
equation~(\ref{kg-0}) and Eq.~(\ref{kgt-gauge}) is that the latter
is a nonstationary Klein-Gordon-type equation unless $\varphi=0$
and $\dot{\vec A}=0$. These conditions are fulfilled only if the
electric field $\vec E=-(\dot{\vec A}+\vec\nabla\varphi)$ vanishes
and the magnetic field $\vec B=\vec\nabla\times\vec A$ is
time-independent.

For a scalar field interacting with an arbitrary stationary
magnetic field we can apply the results of Sections~2 and~3 by
simply replacing the operator $D$ by ${\cal D}$ (and noting that
$\varphi=0$ and $\vec A=\vec A(\vec x)$) or equivalently by
enforcing the minimal coupling prescription: $\vec\nabla\to
\vec\nabla-iq\vec A(\vec x)$.

If either a nonzero electric field or a nonstationary magnetic
field is present, then the transformed field $\chi$ is a
nonstationary Klein-Gordon-type field and one must employ the
quantum mechanics of such fields as outlined in Ref.~\cite{ap}.

\section{Concluding Remarks}

The first-quantized relativistic quantum mechanics for free scalar
fields may be formulated by constructing a genuine Hilbert space
of the solutions of the Klein-Gordon equation. This involves
endowing the solution space of this equation with a
positive-definite inner product. The requirements that this inner
product be well-defined, positive-definite, and relativistically
invariant fix it up to an arbitrary real parameter $a\in(-1,1)$
and an overall trivial coefficient $\kappa\in\R^+$. The resulting
family of inner products, $(\cdot,\cdot)_a$, actually define
unitarily equivalent Hilbert spaces ${\cal H}_a$ and therefore are
physically identical, \cite{ap}. Furthermore, they may be used to
construct a conserved four-vector current density $J_a^\mu$ that
tends to the probability current density for position measurements
in nonrelativistic limit. A by-product of the construction of
$J_a^\mu$ is a manifestly covariant expression for the inner
product $(\cdot,\cdot)_a$.

In view of the unitary-equivalence of ${\cal H}_a$ and
$L^2(\R^3)\oplus L^2(\R^3)$, one can use the ordinary position
operator for a nonrelativistic spin-$1/2$ particle, that acts in
$L^2(\R^3)\oplus L^2(\R^3)$, to define a relativistic position
operator for Klein-Gordon fields $\psi$. This in turn yields a
position basis in which $\psi$ is uniquely determined by a set of
wave functions $f(\epsilon,\vec x)$. In terms of these wave
functions the probability density for the localization of $\psi$
in space takes the same form as in nonrelativistic quantum
mechanics. By expressing $f(\epsilon,\vec x)$ directly in terms of
$\psi$, one obtains a manifestly positive-definite probability
density $\rho_a$. This turns out to coincide with the
Rosenstein-Horwitz probability density \cite{rh}, if one sets
$a=0$ and restricts to the positive-energy Klein-Gordon fields.
One can define a current density ${\cal J}^\mu_a$ whose
zero-component equals $\rho_a$. But ${\cal J}^\mu_a$ is neither
covariant nor conserved.

The probability density $\rho_a={\cal J}^0_a$ may be linked with
the zero-component $J^0_a$ of the conserved current density
$J^\mu_a$ in the sense that their integrals over the whole space
are identical. This in particular ensures the conservation and
frame-independence of the total probability. It also allows for
the interpretation of the continuity equation satisfied by
$J^\mu_a$ as a local manifestation of the (global) conservation
law for the total probability. The latter stems from an underlying
Abelian global gauge-symmetry of the Klein-Gordon equation. The
nature of the corresponding gauge group $G_a$ depends on the
parameter $a$. For rational values of $a$, $G_a=U(1)$; for
irrational values of $a$, $G_a=\R^+$.

The expression for the current densities $J_a^\mu$ obtained for
free Klein-Gordon fields may be easily generalized to scalar
fields interacting with a stationary background magnetic field.
For a more general background electromagnetic field, the scalar
field may be gauge-transformed to a nonstationary
Klein-Gordon-type field. Therefore, in order to understand
first-quantized scalar fields interacting with such an
electromagnetic field one should employ the quantum mechanics of
nonstationary Klein-Gordon-type fields \cite{ap}. This requires a
separate investigation of its own and will be dealt with
elsewhere. Perhaps a more interesting subject of future study is
the local analog of the above-described global $G_a$ gauge
symmetry.

\section*{Acknowledgment}
A.~M.~wishes to acknowledge the support of the Turkish Academy of
Sciences in the framework of the Young Researcher Award Program
(EA-T$\ddot{\rm U}$BA-GEB$\dot{\rm I}$P/2001-1-1).

\section*{Appendices}
\begin{appendix}

\section{Lorentz-Invariance of $D^{-1/2}\dot\psi$}

Consider performing an infinitesimal Lorentz transformation
    \be
    x^\mu\to x'^{\mu} = \Lambda^{\mu}_{~\nu} x^{\nu}, \hspace{1.5cm}
    \Lambda^{\mu}_{~\nu} = \delta^{\mu}_{~\nu} +
    \omega^{\mu}_{~\nu},
    \label{ap-a1}
    \end{equation}
where $\omega^{\mu}_{~\nu}$ are the antisymmetric generators of
the Lorentz transformation \cite{weinberg} and
$|\omega^{\mu}_{~\nu}| \ll 1$. The latter condition means that we
can safely neglect the second and higher order terms in powers of
$\omega^{\mu}_{~\nu}$.

It is not difficult to show using (\ref{ap-a1}) and its inverse
transformation, namely
     \be
     x'^{\mu}\to x^{\mu}=(\Lambda^{-1})^{\mu}_{~\nu} x'^{\nu}=
     x'^{\mu} - \omega^{\mu}_{~\nu} x'^{\nu},
     \label{ap-a2}
     \end{equation}
that
    \be
    \nabla'^2 := \partial'^i\partial'_i =
    \nabla^2 - 2 \omega^\mu_{~i} \partial_\mu\partial^i
     =  \nabla^2 + 2 \omega^\mu_{~0} \partial_\mu\partial^0.
    \label{ap-a3}
    \end{equation}
Here to establish the last equality we have added and subtracted
$2 \omega^\mu_{~0}\partial_\mu\partial^0$ and employed the
antisymmetry of $\omega^\mu_{~\nu}$.

Now, we substitute (\ref{ap-a3}) in $D'=\cum^2 - \nabla'^2$ to
obtain
    \be
    \hat D'^{\alpha} = \hat D^{\alpha} - 2 \alpha \omega^{\mu}_{~0} \hat D^{\alpha-1} \partial_\mu\partial^0~~~~~~~~~~~~~
    \forall \alpha\in\R.
    \label{ap-a4}
    \end{equation}
Moreover, using (\ref{ap-a2}) we have
    \be
    \dot\psi'(x') = \partial'_0 \psi'(x') =
    \dot\psi(x) - \omega^{\mu}_{~0} \partial_\mu \psi(x).
    \label{ap-a5}
    \end{equation}
Finally, in view of (\ref{ap-a4}), (\ref{ap-a5}), and the
Klein-Gordon equation (\ref{kg}), we find
    \bea
    \hat D'^{-1/2} \dot\psi'(x')&=&\left( \hat D^{-1/2} +
    \omega^{\mu}_{~0} \hat D^{-3/2} \partial_\mu
    \partial^0 \right) \left( \dot\psi(x) - \omega^{\mu}_{~0}
    \partial_\mu \psi(x) \right) \nn\\
    &=&\hat D^{-1/2} \dot\psi(x)  - \omega^{\mu}_{~0}
    \left( \hat D^{-3/2} \partial_\mu \ddot\psi(x)
    + \hat D^{-1/2} \partial_\mu\psi(x) \right) \nn\\
    &=&\hat D^{-1/2} \dot\psi(x).\nn
    \eea
Therefore $\hat D^{-1/2}\dot\psi(x)$ is Lorentz-invariant.

\section{Nonrelativistic Limit of $J_a^\mu$}

Let $\psi$ be a Klein-Gordon field and define $\chi:\R^4\to\C$
according to
    \be
    \chi(x^0,\vec x):= e^{i\cum x^0}\psi(x^0,\vec x).
    \label{apb1}
    \end{equation}
Then as it is well-known \cite{greiner}, in the nonrelativistic
limit as $c\to\infty$, $\chi(x)$ satisfies the nonrelativistic
free Schr\"odinger equation:
$\dot\chi=\frac{i}{2\cum}\nabla^2\chi$, and
    \be
    \lim_{c\to\infty}\dot\psi(x)=
    e^{-i\cum x^0}\left\{-i\cum \chi(x) +
    \frac{i}{2\cum} \nabla^2\chi(x)\right\},
    \label{apb2}
    \ee
Furthermore,
    \be
    \lim_{c\to\infty}\hat D^{-1/2}=
    \cum^{-1} +\frac{1}{2}\;\cum^{-3} \nabla^2.
    \label{apb3}
    \ee
Equations~(\ref{psi-c}), (\ref{psi-tilde}), (\ref{apb2}), and
(\ref{apb3}) imply
   \be
   \lim_{c\to\infty} \psi_c=\psi,~~~~~~~~~~~~~~
   \lim_{c\to\infty} \tilde\psi_a=(1+a) \psi.
   \label{apb4}
   \ee
Substituting these relations in (\ref{J-2}), we find
    \be
    \lim_{c\to\infty}J_a^\mu(x)=-\frac{i\kappa(1+a)}{2\cum}
    ~\lim_{c\to\infty}\left[\psi(x)^*
    \stackrel{\leftrightarrow}{\partial^\mu}\psi(x)\right],
    \label{apb5}
    \ee
If we set $\kappa=1/(1+a)$ and $\mu=1, 2, 3$ in this expression,
we obtain (\ref{nonrel-2}). Similarly using (\ref{apb2}) and
(\ref{apb5}), we arrive at (\ref{nonrel-1}). This ends our
demonstration that $J_a^\mu$ has the correct nonrelativistic
limit.

\section{Real and Imaginary Parts of $J_a^\mu$}

Given a free Klein-Gordon field $\psi$, we can use (\ref{psi-pm})
to write
    \[\psi=\psi_++\psi_-,~~~~~~~~~~\psi_c=\psi_+-\psi_-\]
Substituting these relations and (\ref{psi-tilde}) in (\ref{J-2})
and carrying out the necessary calculations, we find the following
expressions for the real and imaginary parts of $J_a^\mu$.
    \bea
    \Re(J_a^\mu)&=&\frac{\kappa}{\cum}\;\Im\left[
    (1+a)\psi_+^*\partial^\mu\psi_+-(1-a)\psi_-^*\partial^\mu\psi_-
    +a(\psi_+^*\partial^\mu\psi_-+\psi_-^*\partial^\mu\psi_+)\right]
    \nn\\
    &=&-\frac{i\kappa}{2\cum}\;\left[
    (1+a)\psi_+^*\stackrel{\leftrightarrow}{\partial^\mu}\psi_+-
    (1-a)\psi_-^*\stackrel{\leftrightarrow}{\partial^\mu}\psi_-
    +2ia\Im(\psi_+^*\stackrel{\leftrightarrow}{\partial^\mu}\psi_-)
    \right],
    \label{a1}\\
    \Im(J_a^\mu)&=&\frac{\kappa}{\cum}\;\Re(
    \psi_+^*\partial^\mu\psi_--\psi_-^*\partial^\mu\psi_+)=
    \frac{\kappa}{\cum}\;\Re(
    \psi_+^*\stackrel{\leftrightarrow}{\partial^\mu}\psi_-).
    \label{a2}
    \eea
Here $\Re$ and $\Im$ stand for the real and imaginary part of
their arguments, respectively.

Note that for a Klein-Gordon field with a definite charge-parity
$\epsilon$, i.e., for a positive- or negative-energy field,
$J_a^\mu$ is real and up to a real coefficient, namely $\pm(1\pm
a)$, coincides with the Klein-Gordon current density. However, due
to the particular sign of this coefficient and the fact that
$|a|<1$, $J_a^0$ is positive-definite for both the positive- and
negative-energy plane-wave solutions of the Klein-Gordon equation.

Another interesting case is that of the real Klein-Gordon fields
for which $\psi_-=\psi_+^*$. Then in view of (\ref{a1}) and
(\ref{a2}), $J^\mu_a=-(i\kappa/\cum)\,\psi_+^*
\stackrel{\leftrightarrow}{\partial^\mu}\psi_+$ which is again
real, but unlike the Klein-Gordon current density it does not
vanish. Note that in this case $J^\mu_a$ is independent of $a$.
This was to be expected, because $a$ enters in the
expression~(\ref{J}) for $J^\mu_a$ as the coefficient of a term
which is essentially the Klein-Gordon current density.

\section{Derivation and Properties of ${\cal J}_a^\mu$}

The derivation of ${\cal J}_a^\mu$ mimics that of $J_a^\mu$.
First, we identify ${\cal J}_a^0$ with $\rho_a$ of
Eq.~(\ref{rho-a-3}). In order to obtain the spatial components of
${\cal J}_a^i$ of ${\cal J}_a^\mu$, we then perform an
infinitesimal Lorentz boost transformation (\ref{boost}) which
yields
   \be
   {\cal J}_a^0(x)\to {\cal J}_a^{'0}(x') =
   {\cal J}_0^\mu(x) - \vec\beta\cdot\vec{\cal J}_a(x).
   \label{boost-calj}
   \end{equation}
Next, we use (\ref{rho-a-3}) to read off the expression for ${\cal
J}_a^{'0}(x')$, namely
    \bea
    {\cal J}_a^{'0}(x')&=&\frac{\kappa}{2\cum}\left\{|\hat D'^{1/4}\psi'(x')|^2+|\hat D'^{1/4}\psi'_c(x')|^2+\right.\nn\\
    &&\vspace{2cm}\left.a\left[(\hat D'^{1/4}\psi'(x'))^* \hat D'^{1/4}\psi'_c(x')+(\hat D'^{1/4}\psi'(x'))
    (\hat D'^{1/4}\psi'_c(x'))^*\right]\right\}.
    \label{calj-1}
    \eea
In view of Eq.~(\ref{q2}) and the fact that $\psi(x)$ and
$\psi_c(x)$ are scalars, we further have
   \bea
   |\hat D'^{1/4}\psi'(x')|^2&=&|\hat D^{1/4}\psi(x)|^2 -
   \frac{1}{2}\vec\beta\cdot\left\{
   (\hat D^{1/4}\psi(x))^* \vec\nabla \hat D^{-3/4}\dot\psi(x) +
   \right.\nn\\
    &&\hspace{.5cm}\left.
   \hat D^{1/4}\psi(x) \vec\nabla (\hat D^{-3/4}\dot\psi(x))^*
   \right\},
   \label{id1}\\
   |\hat D'^{1/4}\psi'_c(x')|^2&=&|\hat D^{1/4}\psi_c(x)|^2 -
   \frac{1}{2}\vec\beta\cdot\left\{(\hat D^{1/4}
   \psi_c(x))^* \vec\nabla \hat D^{-3/4}\dot\psi_c(x) +
   \right.\nn\\
    &&\hspace{.5cm}\left.
   \hat D^{1/4}\psi_c(x) \vec\nabla (\hat D^{-3/4}
   \dot\psi_c(x))^*\right\},
   \label{id2}\\
   (\hat D'^{1/4}\psi'(x'))^* \hat D'^{1/4}\psi'_c(x')&=&
   (\hat D^{1/4}\psi(x))^* \hat D^{1/4}\psi_c(x) -
   \frac{1}{2}\vec\beta\cdot\left\{(\hat D^{1/4}
   \psi(x))^* \vec\nabla \hat D^{-3/4}\dot\psi_c(x) +
    \right.\nn\\
    &&\hspace{.5cm}\left.
   \hat D^{1/4}\psi_c(x)\vec\nabla (\hat D^{-3/4}
   \dot\psi(x))^*\right\},
   \label{id3}\\
   (\hat D'^{1/4}\psi'(x')) (\hat D'^{1/4}\psi'_c(x'))^*&=&
   (\hat D^{1/4}\psi(x)) (\hat D^{1/4}\psi_c(x))^* -
   \frac{1}{2}\vec\beta\cdot\left\{(\hat D^{1/4}\psi_c(x))^*
   \vec\nabla \hat D^{-3/4}\dot\psi(x) +
   \right.\nn\\
   &&\hspace{.5cm}\left.
   \hat D^{1/4}\psi(x) \vec\nabla (\hat D^{-3/4}\dot\psi_c(x))^*
   \right\}.
   \label{id4}
   \eea
Now, substituting (\ref{id1}) - (\ref{id4}) in (\ref{calj-1}) and
making use of (\ref{psi-c}) and (\ref{boost-calj}), we obtain
   \bea
   \vec{\cal J}_a(x)&=&-\frac{i\kappa}{4\cum}\left\{
   (\hat D^{1/4}\psi(x))^* \vec\nabla \hat D^{-1/4}\psi_c(x) -
   \hat D^{1/4}\psi(x) \vec\nabla (\hat D^{-1/4}\psi_c(x))^* +
   \right. \nn\\
   &&\vspace{2cm}\left.(\hat D^{1/4}\psi_c(x))^* \vec\nabla
   \hat D^{-1/4}\psi(x) -
   \hat D^{1/4}\psi_c(x) \vec\nabla (\hat D^{-1/4}\psi(x))^* +
   \right. \nn\\
   &&\vspace{2cm}\left.a\left[(\hat D^{1/4}\psi(x))^* \vec\nabla
   \hat D^{-1/4}\psi(x) -
   \hat D^{1/4}\psi(x) \vec\nabla (\hat D^{-1/4}\psi(x))^*
   \right.\right.\nn\\
   &&\vspace{2cm}\left.\left.(\hat D^{1/4}\psi_c(x))^* \vec\nabla
    \hat D^{-1/4}\psi_c(x) -
   \hat D^{1/4}\psi_c(x) \vec\nabla (\hat D^{-1/4}\psi_c(x))^*
   \right]\right\}.
   \label{cal-vec-j}
   \eea
This relation suggests
   \bea
   {\cal J}_a^\mu(x)&=&-\frac{i\kappa}{4\cum}
   \left\{(\hat D^{1/4}\psi(x))^* \partial^\mu
   \hat D^{-1/4}\psi_c(x) -
   \hat D^{1/4}\psi(x) \partial^\mu (\hat D^{-1/4}
   \psi_c(x))^* + \right. \nn\\
   &&\vspace{2cm}\left.(\hat D^{1/4}\psi_c(x))^*
   \partial^\mu \hat D^{-1/4}\psi(x) -
   \hat D^{1/4}\psi_c(x) \partial^\mu (\hat D^{-1/4}
   \psi(x))^* + \right. \nn\\
   &&\vspace{2cm}\left.a\left[(\hat D^{1/4}\psi(x))^*
   \partial^\mu \hat D^{-1/4}\psi(x) -
   \hat D^{1/4}\psi(x) \partial^\mu (\hat D^{-1/4}\psi(x))^*
   \right.\right.\nn\\
   &&\vspace{2cm}\left.\left.(\hat D^{1/4}\psi_c(x))^*
   \partial^\mu \hat D^{-1/4}\psi_c(x) -
   \hat D^{1/4}\psi_c(x) \partial^\mu (\hat D^{-1/4}
   \psi_c(x))^*\right]\right\},
   \label{calj-2}
   \eea
which we can also write as
   \bea
   {\cal J}_a^\mu(x)&=&\frac{\kappa}{2\cum}\Im\left\{
   (\hat D^{1/4}\psi(x))^* \partial^\mu \hat D^{-1/4}\psi_c(x) -
   (\hat D^{1/4}\psi_c(x)) \partial^\mu (\hat D^{-1/4}\psi(x))^* +
   \right.\nn\\
   &&\vspace{2cm}\left.a\left[(\hat D^{1/4}\psi(x))^* \partial^\mu
   \hat D^{-1/4}\psi(x) -
   (\hat D^{1/4}\psi_c(x)) \partial^\mu (\hat D^{-1/4}
   \psi_c(x))^*\right]\right\}.
   \label{calj-3}
   \eea
Using Klein-Gordon equation, we can easily check that the
expression for ${\cal J}_a^0(x)=\rho_a(x)$ obtained by setting
$\mu=0$ in (\ref{calj-3}) agrees with the one given in
(\ref{rho-a-3}).

Next, we explore the classical limit of the probability current
density (\ref{calj-3}). Following the treatment of Appendix~B, we
first derive
    \[\lim_{c\to\infty}\hat D^{1/4}=\cum^{1/2} -
    \frac{1}{4}\;\cum^{-3/2} \nabla^2,~~~~~~~~
    \lim_{c\to\infty}\hat D^{-1/4}=\cum^{-1/2} +
    \frac{1}{4}\;\cum^{-5/2} \nabla^2.\]
Substituting these relations in (\ref{calj-3}), we find
    \[\lim_{c\to\infty}{\cal J}_a^\mu=
    -\frac{i\kappa(1+a)}{2\cum}\lim_{c\to\infty}\left[
    \psi(x)^* \stackrel{\leftrightarrow}{\partial^\mu} \psi(x)
    \right].\]
Now, setting $\kappa=1/(1+a)$, considering $\mu=0$ and $\mu\neq 0$
separately, and making use of (\ref{apb1}) and (\ref{apb2}), we
obtain (\ref{class3}):
    \[\lim_{c\to\infty} {\cal J}_a^0(x^0,\vec x)=\varrho
    (x^0,\vec x),~~~~~~~~~
    \lim_{c\to\infty} {\cal \vec J}_a(x^0,\vec x)=\frac{1}{c}\;
        \vec j(x^0,\vec x),\]
where $\varrho$ and $\vec j$ are the classical scalar and current
probability densities given by (\ref{class1}) and (\ref{class2}),
respectively.

If we confine our attention to the positive-energy Klein-Gordon
fields (for which $\psi_c=\psi$) and set $a=0$, the expression
(\ref{calj-2}) coincides with the Rosenstein-Horwitz's current
\cite{rh}:
   \be
   {\cal J}_{\rm RH}^\mu(x)=-\frac{i\kappa}{2\cum}\left\{
   (\hat D^{1/4}\psi(x))^* \partial^\mu (\hat D^{-1/4}\psi(x)) -
   (\hat D^{1/4}\psi(x)) \partial^\mu (\hat D^{-1/4}\psi(x))^*
   \right\}.
   \label{rh-current}
   \end{equation}

As we show below, in general $\partial_\mu{\cal J}_a^\mu \neq 0$.
Hence ${\cal J}_a^\mu$ is not a conserved current density. This
was noticed by Rosenstein and Horwitz \cite{rh} for the
probability current (\ref{rh-current}). A more dramatic result
that seems to be missed by these authors is that ${\cal J}_{\rm
RH}^\mu$ is not even a four-vector field. The same holds for
${\cal J}_a^\mu$. This can be most conveniently shown by computing
${\cal J}_a^\mu$ for a superposition of a pair of positive-energy
plane-wave Klein-Gordon fields:
    \be
    \psi(x)=c_1 e^{i k_1\cdot x} + c_2 e^{i k_2\cdot x} =
    c_1 e^{-i\omega_1 x^0} e^{i \vec k_1\cdot \vec x}
    + c_2 e^{-i\omega_2 x^0} e^{i \vec k_2\cdot \vec x},
    \label{apd-8}
    \end{equation}
where for $\ell\in\{1,2\}$, $c_\ell\in\C-\{0\}$, $k_\ell$ is a
constant four-vector, $k_\ell\cdot x:=(k_\ell)_\mu x^\mu$, and
$\omega_\ell:=\sqrt{\vec k_\ell^2+\cum^2}$. Note that for this
choice of the field we have $\psi_c=\psi$ and ${\cal
J}_0^\mu={\cal J}_{\rm RH}^\mu$. Hence, in view of (\ref{apd-8})
and (\ref{rh-current}),
    \be
    {\cal J}_0^\mu(x)={\cal J}_{\rm RH}^\mu(x)=\frac{\kappa}{\cum}
    \left\{|c_1|^2 k_1^{\mu} +
    |c_2|^2 k_2^{\mu} + \Re\left( c_1 c_2^* e^{i (k_1 - k_2).x}
    \right) K^{\mu}\right\},
    \label{apd-9}
    \end{equation}
where
    \be
    K^\mu = \sqrt{\frac{\omega_2}{\omega_1}}\;
    k_1^\mu + \sqrt{\frac{\omega_1}{\omega_2}}\; k_2^\mu.
    \label{apd-3}
    \end{equation}
Moreover, setting $\psi_c=\psi$ in (\ref{calj-3}), we find
    \be
    {\cal J}_a^\mu(x)=(1+a){\cal J}_0^\mu(x).
    \label{zzz}
    \ee
A direct implication of Eqs.~(\ref{apd-9}) and (\ref{zzz}) is that
${\cal J}_a^\mu$ is a vector field if and only if $K^\mu$ is a
four-vector. But as we show next the latter fails to be the case.

Suppose (by contradiction) that $K^\mu$ is a four-vector, then
$K_\mu K^\mu$ must be a scalar. It is not difficult to show that
    \be
    K_\mu K^\mu = 2 k_1\cdot k_2 -\cum^2\left(
    \frac{\omega_2}{\omega_1}+
    \frac{\omega_1}{\omega_2}\right),
    \label{apd-4}
    \end{equation}
where we have made use of $k_\ell\cdot k_\ell=-\cum^2$. For
$\omega_1\neq\omega_2$, the term multiplying $\cum^2$ on the
right-hand side of (\ref{apd-4}) fails to be a scalar. This shows
that $K_\mu K^\mu$ is not a scalar; $K^\mu$ is not a four-vector;
and in general ${\cal J}_a^\mu$ is not a vector field.

Next, we wish to point out that computing the current density
$J_a^\mu$ for the field~(\ref{apd-8}) we find the following
manifestly covariant expression.
    \be
    J_a^\mu(x) = \frac{\kappa (1+a)}{\cum}\left\{|c_1|^2 k_1^{\mu} +
    |c_2|^2 k_2^{\mu} + \Re
    \left(c_1 c_2^* e^{i (k_1 - k_2).x}\right)
    \left(k_1^\mu + k_2^\mu\right)\right\}.
    \label{xzx}
    \ee

Having obtained the explicit form of both ${\cal J}^\mu_a$ and
$J^\mu_a$ for the field (\ref{apd-8}), we can easily check their
conservation property. A simple calculation shows that
    \bea
    \partial_\mu {\cal J}^\mu_a(x)&=& (\cum^2+k_1\cdot k_2)
    \left(\sqrt{\frac{\omega_1}{\omega_2}}-
    \sqrt{\frac{\omega_2}{\omega_1}}\right)\:{\cal F}(x),
    \label{zz1}\\
    \partial_\mu J^\mu_a(x)&=&[(k_1-k_2)\cdot (k_1+k_2)]
    \;{\cal F}(x)= 0,
    \label{zz2}
    \eea
where
    \[{\cal F}(x):= -\frac{\kappa (1+a)}{\cum}\;
    \Im\left[c_1 c_2^* e^{i (k_1 - k_2).x}\right],\]
and we have made use of (\ref{zzz}), (\ref{apd-9}), (\ref{apd-3}),
and (\ref{xzx}) and the fact that the term in the square bracket
on the right hand side of (\ref{zz2}) vanishes identically by
virtue of $k_\ell\cdot k_\ell=-\cum^2$. According to
Eq.~(\ref{zz1}), for $\omega_1\neq\omega_2$, $\partial_\mu {\cal
J}^\mu_a(x)\neq 0$. Therefore, unlike $J_a^\mu$, the probability
current density ${\cal J}_a^\mu$ fails to be conserved.

\end{appendix}

\ed
\newpage

\begin{thebibliography}{99}
\bibitem{holstein} B.~R.~Holstein, {\em Topics in Advanced Quantum
Mechanics} (Addison-Wesley, Redwood City, CA, 1992)
\bibitem{ghose} P.~Ghose, M.~K.~Samal, and A.~Datta, Phys.\
Lett.~A {\bf 315}, 23 (2003).
\bibitem{peierls} R.~Peierls, {\em Surprises in Theoretical
Physics} (Princeton University Press, Princeton, 1979).
\bibitem{greiner} W.\ Greiner, {\em Relativistic Quantum Mechanics}
(Springer, Berlin, 1994).
\bibitem{wald-gr} R.~M.~Wald, {\em General Relativity} (Chicago
University Press, Chicago, 1984).
\bibitem{causality} G.~C.~Hegerfeldt, Phys.\ Rev.~D {\bf 10}, 3320
(1974); ibid {\bf 22}, 377 (1980); and Phys.\ Rev.\ Lett.~{\bf
54}, 2395 (1985).
\bibitem{zuben} F.~S.~G.~Von~Zuben, J.\ Math.\ Phys.\ {\bf 41},
6093 (2000).
\bibitem{barat-kimball} N.~Barat and J.~C.~Kimball, Phys.\ Lett.~A
{\bf 308}, 110 (2003).
\bibitem{ryder} L.~Ryder, {\em Quantum Field Theory} (Cambridge
University Press, Cambridge, 1996).
\bibitem{bryce-67} B.\ S.\ DeWitt, Phys.\ Rev.\ {\bf 160}, 1113
(1967).
\bibitem{qg-review} K.\ Kuch\'ar, in {\em Proceedings of the 4th Canadian Conference
on Relativity and Relativistic Astrophysics}, edited by
G.~Kunstatter, D.~Vincent, and J.~Williams (World Scientific,
Singapore, 1992);\\
C.\ J.\ Isham, in {\em Integrable Systems, Quantum Groups, and
Quantum Field Theories,} edited by L.\ A.\ Ibort and M.\ A.\
Ropdriguez (Kluwer, Dordrecht, 1993);\\
S.~Carlip, Rep.\ Prog.\ Phys.\ {\bf 64}, 885 (2001).
\bibitem{indefinite-m} J.\ Bogn\'ar, {\em Indefinite Inner Product
Spaces} (Springer, Berlin, 1974);\\
T.\ Ya.\ Azizov and I.\ S.\ Iokhvidov, {\em Linear Operators in
Spaces with Indefinite Metric} (Wiley, Chichester, 1989).
\bibitem{indefinite} W.~Pauli, Rev.\  Mod.\ Phys., {\bf
15}, 175 (1943); \\
S.~N.~Gupta, Proc.\ Phys.\ Soc.\ London {\bf 63}, 681 (1950);\\
K.~Bleuler, Helv.\ Phys.\ Acta {\bf 23}, 567 (1950);\\
E.~C.~G.~Sudarshan, Phys.\ Rev.\ {\bf 123}, 2183 (1961);\\
T.~D.~Lee and G.~C.~Wick, Nucl.\ Phys~B {\bf 9}, 209 (1969);\\
A.~Mostafazadeh, Czech J.~Phys.~{\bf 53}, 1079 (2003); preprint:
quant-ph/0308028.
\bibitem{reed-simon}  M.\ Reed and B.~Simon, {\em Functional
Analysis,} vol.\ I (Academic Press, San Diego, 1980).
\bibitem{wald-qft} B.~S.~DeWitt, Phys.\ Rep.~{\bf 19}, 295
(1975);\\
R.~M.~Wald, {\em Quantum Field Theory in Curved
Spacetime and Black Hole Thermodynamics} (Chicago University
Press, Chicago, 1994).
\bibitem{cqg} A.~Mostafazadeh, Class.\ Quantum Grav.\ {\bf 20}, 155
(2003).
\bibitem{ap} A.~Mostafazadeh, `Quantum Mechanics of
Klein-Gordon-Type Fields and Quantum Cosmology,' preprint:
gr-qc/0306003, Ann.~Phys.~(N.Y.), to appear.
\bibitem{p57} A.~Mostafazadeh,`Generalized PT-, C-, and CPT-Symmetries, Position Operators, and
Localized States of Klein-Gordon Fields,' preprint:
quant-ph/0307059.
\bibitem{ph} A.\ Mostafazadeh, J.\ Math.\ Phys.\ {\bf 43},
205 (2002); ibid {\bf 43}, 2814 (2002); ibid {\bf 43}, 3944
(2002); ibid {\bf 44}, 974 (2003); and Nucl.\ Phys.~B {\bf 640},
419 (2002).
\bibitem{halliwell-ortiz} J.~J.~Halliwell and M.~E.~Ortiz, Phys.\
Rev.~D {\bf 48}, 748 (1993).
\bibitem{woodard} P.\ P.\ Woodard, Class.\ Quantum.\ Grav.\
{\bf 10}, 483 (1993).
\bibitem{hartle-marolf} J.~B.~Hartle and D.~Marolf, Phys.~Rev.~D
{\bf 56}, 6247 (1997).
\bibitem{halliwell-thorwart} J.~J.~Halliwell and J.~Thorwart,
Phys.~Rev.~D {\bf 64}, 124018 (2001).
\bibitem{induced} A.~Ashtekar, J.~Lewandowski, D.~Marolf,
J.~Mour\~ao, and T.~Thiemann, J.~Math.\ Phys.\ {\bf 36}, 6456
(1995);\\
D.\ Marolf, Class.\ Quantum Grav.\ {\bf 12}, 1199 (1995);\\
D.~Giulini and D.\ Marolf, Class.\ Quantum Grav.\ {\bf 16}, 2479
(1999); ibid 2489 (1999).
\bibitem{embacher} F.~Embacher, Hadronic J.~{\bf 21}, 337
(1998);\\
D.\ Marolf, `Group Averaging and Refined Algebraic Quantization:
Where are we now?' preprint: gr-qc/0011112.
\bibitem{rh} B.~Rosenstein and L.~P.~Horwitz, J.\ Phys.\ A:
Math.\ Gen.\ {\bf 18}, 2115 (1985).
\bibitem{holland} P.~Holland, `Uniqueness of Conserved Currents in
Quantum Mechanics,' preprint: quant-ph/0305175.
\bibitem{kato} T.~Kato, {\em Perturbation Theory for Linear
Operators} (Springer, Berlin, 1995).
\bibitem{newton-wigner} T.~D.~Newton and E.~P.~Wigner, Rev.\ Mod.\
Phys.\ {\bf 21}, 400 (1949).
\bibitem{foldy} L.~L.~Foldy, Phys.\ Rev.\ {\bf 102}, 568 (1956).
\bibitem{itzykson-zuber} C.~Itzykson and J.-B.~Zuber, {\em Quantum
Field Theory} (McGraw-Hill, New York, 1980).
\bibitem{weinberg} S.~Weinberg, {\em The Quantum Theory of Fields}
(Cambridge University Press, Cambridge, 1995).
\bibitem{brocker-dieck} T.~Br\"ocker and T.~T.~Dieck, {\em
Representations of Compact Lie Groups,} (Springer, Berlin, 1985).

\end{thebibliography}
